%% file: main.tex
\documentclass[a4paper, amsfonts, amssymb, amsmath, reprint, showkeys, nofootinbib, twoside]{revtex4-1}
\usepackage[english]{babel}
\usepackage[utf8]{inputenc}
\usepackage[colorinlistoftodos, color=green!40, prependcaption]{todonotes}
\input{preamble}

\newcommand{\figref}{Fig.\,\ref}
\newcommand{\secref}{Sec.\,\ref}
\newcommand{\tabref}{Tab.\,\ref}
\newcommand{\Eqref}{Eq.\,\ref}

\begin{document}
\title{Carrier diffusion in semiconductor nanoscale resonators}

\author{Marco Saldutti, Yi Yu, George Kountouris, Philip Trøst Kristensen and Jesper Mørk}
    \email[Correspondence email address: ]{jesm@dtu.dk}
    \affiliation{DTU Electro, Technical University of Denmark, DK-2800 Kgs. Lyngby, Denmark}
    \affiliation{NanoPhoton - Center for Nanophotonics, Technical University of Denmark, DK-2800 Kgs. Lyngby, Denmark}

\date{\today} 

\begin{abstract}
It is shown that semiconductor nanoscale resonators with extreme dielectric confinement accelerate the diffusion of electron-hole pairs excited by nonlinear absorption. These novel cavity designs may lead to optical switches with superior modulation speeds compared to conventional geometries. The response function of the effective carrier density is computed by an efficient eigenmode expansion technique. A few eigenmodes of the diffusion equation conveniently capture the long-timescale carrier decay rate, which is advantageous compared to time-domain simulations. Notably, the eigenmode approach elucidates the contribution to carrier diffusion of the in-plane and out-of-plane cavity geometry, which may guide future designs. 
\end{abstract}


\maketitle

\input{sections/section01.tex}
\input{sections/section02.tex}
\input{sections/section03.tex}
\input{sections/section04.tex}

\input{sections/section05.tex}
\input{sections/section06.tex}


\bibliography{main.bib}

\end{document}

%% file: preamble.tex
\usepackage{amsthm}
\usepackage{mathtools}
\usepackage{physics}
\usepackage{xcolor}
\usepackage{graphicx}
\usepackage[left=23mm,right=13mm,top=35mm,columnsep=15pt]{geometry} 
\usepackage{adjustbox}
\usepackage{placeins}
\usepackage[T1]{fontenc}
\usepackage{lipsum}
\usepackage{csquotes}

%% file: sections/section01.tex
\section{Introduction} \label{sec:introduction}

The ability to control and route optical signals with light itself - namely, all-optical switching \cite{Chai_AdvOptMat_2017,Taghinejad_ACSPhot_2019,Ono_NatPhot_2020,Guo_NatPhot_2022} - is an essential functionality of photonic integrated circuits. In particular, optical switches based on semiconductor nanoscale resonators \cite{Husko_APL_2009,Nozaki_NatPhot_2010,Yu_OptExpress_2013,Bazin_APL_2014,Moille_LPR_2016,Colman_PRL_2016,Saudan_OptExpress_2022} offer high speed, energy efficiency, and reduced footprint, with record-low \cite{Nozaki_NatPhot_2010} switching energies. In these devices, a control signal excites electron-hole pairs (carriers) via two-photon absorption (as well as linear absorption in \cite{Nozaki_NatPhot_2010}), and the free carrier-induced dispersion tunes the cavity refractive index. The cavity resonance shifts, thus blocking or letting through the probe signal encoding the information to be transmitted. 

\begin{figure*}
\centering
\includegraphics[width=1.95\columnwidth]{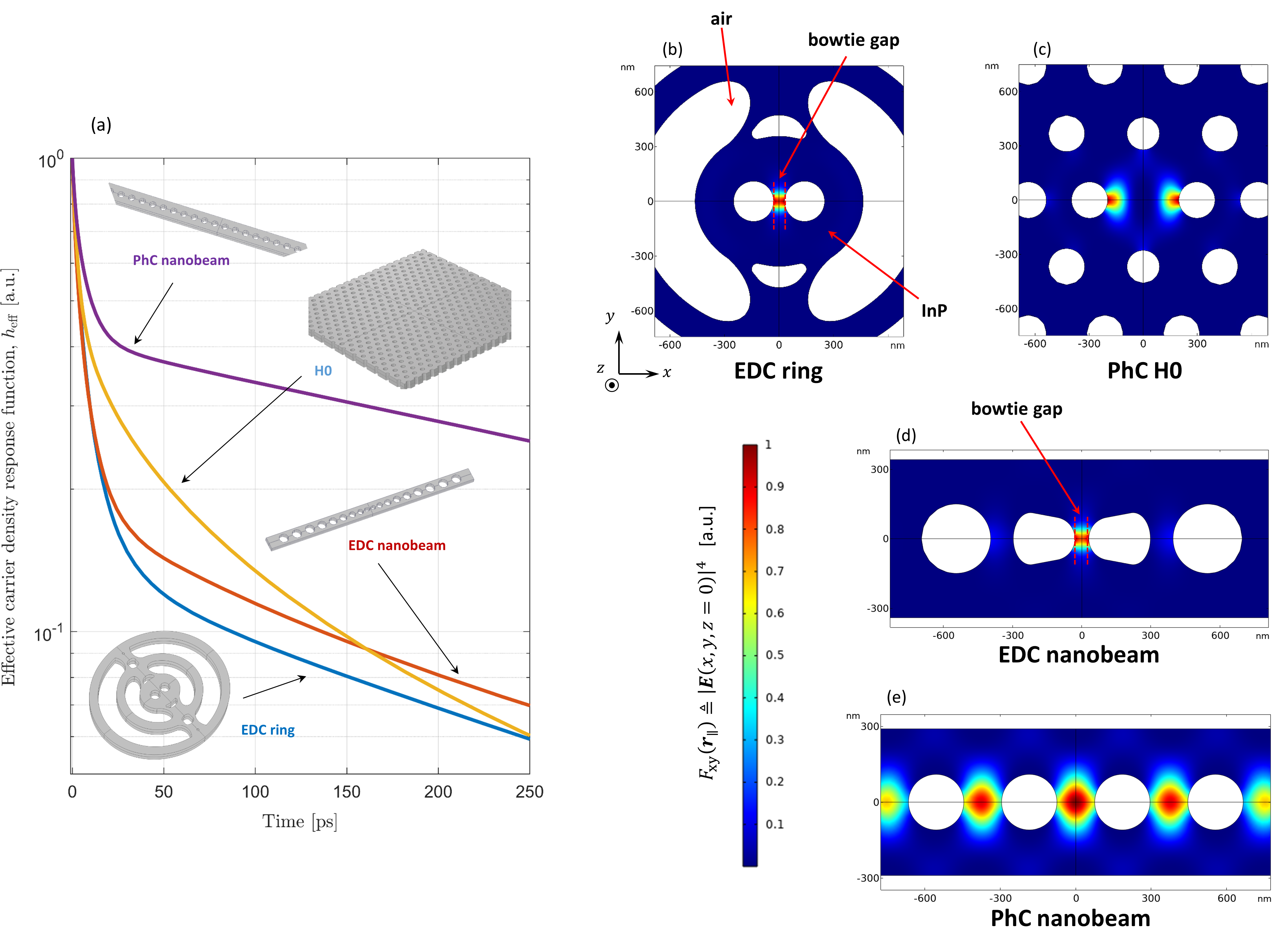}
\caption{(a) Response function of the mode-averaged (or effective) carrier density for a ring cavity (blue) with extreme dielectric confinement (EDC), an EDC nanobeam cavity (red), a photonic crystal (PhC) H0 cavity (yellow) and a PhC nanobeam cavity (purple). The cavity parameters are summarized in \tabref{tab:sec:SEC1:TAB1}. The in-plane excitation profile of the carrier density due to two-photon absorption in the various cavities is also shown, as indicated in the following: (b) EDC ring, (c) PhC H0, (d) EDC nanobeam, and (e) PhC nanobeam. The response function is obtained from full three-dimensional simulations. The surface recombination velocity at the boundaries between the semiconductor material and surrounding air is $10^4\,\mathrm{cm/s}$. The designs in (b), (c), and (e) are inspired by \cite{Kountouris-OptExress-2022}, \cite{Yu_OptExpress_2013} and \cite{Bazin_JLWT_2014}, respectively.}
\label{fig:sec:SEC1:FIG1-FIG2} 
\end{figure*}
However, the modulation speed is limited by the diffusion and recombination time of the excited carriers, and data transmission rates exceeding a few tens of Gbit/s are yet to be demonstrated \cite{Bekele_LaserPhotRev_2019}. Previous works \cite{Nozaki_NatPhot_2010,Yu_OptExpress_2013,Moille_PRA_2016} on photonic crystal (PhC) cavities \cite{Saldutti_Nanomat_2021} have shown that tight field confinement reduces the time it takes for the carriers to diffuse out of the effective mode area of interest. Here, we show that new cavity designs comprising a bowtie and featuring deep sub-wavelength optical confinement \cite{Hu_ACSPhot_2016,Choi_PRL_2017,Wang_APL_2018,Albrechtsen_NatComm_2022} (so-called extreme dielectric confinement, EDC) further speed up the diffusion of carriers. The large quality factor (Q-factor) and enhanced carrier diffusion may lead to energy-efficient optical switches with superior modulation speeds. 

The diffusion and recombination of carriers are often modeled by rate equations, with different numbers of time constants \cite{Johnson_OE_2006, Heuck_APL_2013,Yu_OptExpress_2013}. The time constants are usually determined by fitting to experiments or space- and time-domain simulations of the ambipolar diffusion equation \cite{Tanabe_JLWT_2008,Nozaki_NatPhot_2010,Yu_OptExpress_2013}. However, it has been pointed out that even a multi-exponential decay may not be entirely satisfactory and that a Green's function formalism generally provides better accuracy \cite{Moille_PRA_2016}. The Green's function of the mode-averaged (or effective) carrier density reflects the shift in the cavity resonance due to an impulse excitation of carriers in the time domain. \figref{fig:sec:SEC1:FIG1-FIG2}a shows the Green's function - from now on, \textit{response function} - of the effective carrier density for the cavities considered in this article: EDC ring (blue), EDC nanobeam (red), PhC H0 (yellow), and PhC nanobeam (purple). The cavity parameters are summarized in \tabref{tab:sec:SEC1:TAB1}. The in-plane excitation profile of the carrier density due to two-photon absorption is illustrated in \figref{fig:sec:SEC1:FIG1-FIG2}b, c, d and e. 

In \tabref{tab:sec:SEC1:TAB1}, in addition to the geometrical parameters (in-plane footprint and cavity thickness), we include the quality factor (Q-factor), as well as the two-photon absorption (TPA) and free-carrier absorption (FCA) mode volume \cite{Saldutti_Nanomat_2021}. These parameters are computed from three-dimensional simulations of Maxwell's equations in the frequency domain with radiation boundary conditions \cite{Kountouris-OptExress-2022}. In semiconductor nanocavities, the intensity of the carrier generation rate due to TPA and the induced nonlinear losses scale with the inverse of the nonlinear mode volumes, as detailed in \cite{Yu_OptExpress_2013}. Compared to the PhC cavities, the Q-factor of the EDC cavities in \tabref{tab:sec:SEC1:TAB1} is relatively low. However, the in-plane footprint of the PhC cavities is much larger. The Q-factor of the EDC cavities could be increased by adding air rings or air holes without altering the geometry in the bowtie proximity \cite{Wang_APL_2018}. We emphasize that plasmonic resonators also feature deep sub-wavelength optical confinement \cite{Schuller_NatMat_2010}. However, the Q-factor of the EDC cavities in \tabref{tab:sec:SEC1:TAB1} is already much larger than offered by plasmonic cavities \cite{Wang_PRL_2006,Khurgin_NatNanotech_2015}.   

The response function in \figref{fig:sec:SEC1:FIG1-FIG2}a is obtained from full three-dimensional simulations of the ambipolar diffusion equation in space and time. Surface recombination at the boundaries between the semiconductor material and surrounding air is also included. Compared to more conventional geometries, the electric field in EDC cavities is tightly localized to a hot spot, which accelerates the effective carrier density decay rate. As discussed later, further improvements are feasible by scaling down the bowtie gap.

However, depending on the timescale of interest, time-domain simulations may be computationally demanding. Furthermore, three-dimensional simulations tend to obscure the impact of the cavity geometry in the $xy$-plane (in-plane) and along the growth direction, $z$ (out-of-plane). The $x$-, $y$- and $z$-directions are indicated in \figref{fig:sec:SEC1:FIG1-FIG2}b. 

In this article, we present an alternative eigenmode expansion technique to calculate the response function of the effective carrier density. This eigenmode approach efficiently provides the long-timescale decay rate with only a few eigenmodes of the ambipolar diffusion equation. Notably, the eigenmode approach singles out the contributions of the in-plane and out-of-plane diffusion dynamics, thus offering new insights, that will guide future cavity designs, e.g., by topology optimization \cite{Jensen-Sigmnud_LPR_2011,Molesky_NatPhot_2018}.    

The article is organized as follows. \secref{sec: diffusion model} introduces the carrier diffusion model. In particular, the approximations leading from the drift-diffusion model to the ambipolar diffusion regime are illustrated step by step, and the limitations are discussed, including the impact of surface recombination. The eigenmode expansion and response function formalism are presented in \secref{sec: effective carrier density - eig exp}. The eigenmode approach is then applied in \secref{sec: in-plane diffusion} and \secref{sec: out-of-plane diffusion}, respectively, to analyze the in-plane and out-of-plane diffusion dynamics in detail, with an overview of the cavities in \tabref{tab:sec:SEC1:TAB1}. Finally, \secref{sec: conclusions} draws the main conclusions.  

\begin{table*}[ht]
\footnotesize	
\centering
\begin{tabular}{p{0.15\linewidth}p{0.10\linewidth}p{0.11\linewidth}p{0.12\linewidth}p{0.20\linewidth}p{0.12\linewidth}}
\hline
Cavity & Q-factor & $V_{\mathrm{TPA}}\,\mathrm{[\lambda_0/n]^3}$ & $V_{\mathrm{FCA}}\,\mathrm{[\lambda_0/n]^3}$ & In-plane footprint & \begin{tabular}{@{}c@{}}Cavity\\thickness $\mathrm{[nm]}$\end{tabular}\\
\hline
EDC ring & $2\times10^3$ & 0.5 & 0.3 & \begin{tabular}{@{}c@{}}$2a\!=\!3.2\,\mu\mathrm{m}$, $2b\!=\!2.8\,\mu\mathrm{m}$\\(ellipse axes)\end{tabular} & 240\\
EDC nanobeam & $8\times10^3$ & 0.8 & 0.5 & $11.5\,\mu\mathrm{m}\times683.4\,\mathrm{nm}$ & 250\\
PhC H0 & $3\times10^4$ & 1.7 & 1 & $8.5\,\mu\mathrm{m}\times8.8\,\mu\mathrm{m}$ & 340\\
PhC nanobeam & $3\times10^6$ & 3 & 2 & $21.5\,\mu\mathrm{m}\times580\,\mathrm{nm}$ & 250\\
\hline
\end{tabular}
\caption{\label{tab:sec:SEC1:TAB1} Parameters representative of the photonic cavities in \figref{fig:sec:SEC1:FIG1-FIG2} with resonant wavelength $\lambda_0\approx1550\,\mathrm{nm}$: quality factor (Q-factor), two-photon absorption (TPA) and free-carrier absorption (FCA) mode volume \cite{Saldutti_Nanomat_2021}, size along the $x$- and $y$-direction (in-plane footprint) and size along the $z$-direction (cavity thickness). The outer in-plane perimeter of the EDC ring cavity is elliptical, and the ellipse axes are reported. The EDC cavities have a bowtie gap of $60\,\mathrm{nm}$. Further details on the in-plane geometries are in \figref{fig:sec:SEC1:FIG1-FIG2}. The semiconductor material is Indium Phosphide (InP), with refractive index $n=3.17$.}  
\end{table*}

%% file: sections/section02.tex
\section{Carrier diffusion model: from drift-diffusion to ambipolar diffusion} \label{sec: diffusion model}

The drift-diffusion equations are pivotal for understanding fundamental characteristics of electronic and photonic devices, including transistors \cite{Sze-book3rdEd-2006}, photodetectors \cite{Liu-book-2005} and solar cells \cite{Nelson-book-2003}, and have also been applied to semiconductor lasers \cite{Tessler_JQE_1993,Gready_JSTQE_2013,Tibaldi_JSTQE_2019,Saldutti_PhotRes_2020}, for cases where simpler descriptions based on rate equations may be inadequate. For instance, effects such as the quenching of the ground-state power in the presence of dual-state lasing, or the impact of p-type modulation doping on the lasing threshold are naturally explained \cite{Saldutti_PhotRes_2020} within a drift-diffusion picture.

The optical switching characteristics of semiconductor nanocavities are well understood in a modeling framework that combines temporal coupled-mode theory with the ambipolar regime of the drift-diffusion equations \cite{Nozaki_NatPhot_2010,Yu_OptExpress_2013,Moille_LPR_2016}. However, the literature is not always clear on the approximations involved, and conflicting descriptions flourish, especially regarding the recombination terms. In the following, we derive the ambipolar diffusion equation from the classical drift-diffusion equations, outlining the main approximations and limitations.

The drift-diffusion model \cite{Vasileska_book_2010} consists of the continuity equations for electrons and holes, typically coupled to Poisson's equation. Carrier heating effects \cite{Mark_APL_1992} are ignored, assuming that the excited electrons and holes have already relaxed to their quasi-equilibrium distributions, described by quasi-Fermi levels. The approximation is appropriate on timescales longer than about $1\,\mathrm{ps}$ \cite{Moille_PRA_2016}, to which the conclusions of this article are therefore restricted. The continuity equations read
\begin{subequations}
\begin{align}
\frac{\partial n}{\partial t} & = \frac{1}{q}\nabla\cdot\mathbf{J}_n - U_n + G_n\label{eq:App0:DD-eqs-1}\\
\frac{\partial p}{\partial t} & = -\frac{1}{q}\nabla\cdot\mathbf{J}_p - U_p + G_p \label{eq:App0:DD-eqs-2}
\end{align}
\end{subequations}
Here, $n$ and $p$ are the densities per unit volume of electrons and holes, respectively, whereas $q$ is the electron charge. The electron ($\mathbf{J}_n$) and hole ($\mathbf{J}_p$) current density per unit area reflects the motion of electrons and holes and accounts for both drift and diffusion. 

The generation rates per unit volume and unit time describe the excitation of excess electrons ($G_n$) and holes ($G_p$) with respect to thermal equilibrium. We assume optical excitation, which is fast compared to the timescales of the other processes. Therefore, electrons and holes are generated in pairs, leading to the same generation rate, $G_n = G_p = G$.

The recombination rates per unit volume and unit time of electrons ($U_n$) and holes ($U_p$) account for trap-assisted recombination in the bulk semiconductor material, as well as radiative and Auger recombination. At carrier densities around $10^{15}\sim10^{18}\,\mathrm{cm^{-3}}$, as relevant \cite{Borghi_OptExpress_2021} for optical switching applications, trap-assisted recombination usually dominates over radiative and Auger recombination \cite{Moille_PRA_2016,Moille_OptLett_2017}, which are characterized by longer lifetimes.

Trap-assisted recombination deserves special consideration. Strictly speaking, it is only under steady-state conditions that the general theory \cite{Shockley_PhysRev_1952} of electron and hole trap-assisted recombination reduces to the well-known Shockley-Read-Hall (SRH) recombination rate \cite{Shockley_PhysRev_1952,Hall_PhysRev_1952,Pierret_book_2002}
\begin{equation}
\label{eq:App0:SRH-rate}
U_n = U_p = U = \frac{np - n_i^2}{\tau^{\mathrm{SRH}}_{n}(p+p_1) + \tau^{\mathrm{SRH}}_{p}(n+n_1)}
\end{equation}
with $n_1$ and $p_1$ given by
\begin{subequations}
\begin{align}
n_1 & = n_i\,e^{\frac{E_t-E_{\mathrm{F}_i}}{k_BT}}\label{eq:App0:n1}\\
p_1 & = n_i\,e^{\frac{E_{\mathrm{F}_i}-E_t}{k_BT}} \label{eq:App0:p1}
\end{align}
\end{subequations}
Here, $n_i$ is the electron and hole intrinsic concentration, $E_{\mathrm{F}_i}$ is the intrinsic Fermi level and $E_t$ is the trap energy level, including the trap degeneracy factor \cite{Pierret_book_2002}. A single trap level dominates trap-assisted bulk recombination \cite{Pierret_book_2002}. The Boltzmann constant and temperature are denoted by $k_B$ and $T$, respectively. In \Eqref{eq:App0:SRH-rate}, the probability per unit time of an electron (hole) being trapped when the traps are all empty is $1/\tau^{\mathrm{SRH}}_{n}$ ($1/\tau^{\mathrm{SRH}}_{p}$). This probability is inversely proportional to the density of traps per unit volume.   

In general, electrons and holes feature different capture cross-sections, which leads to $U_n\neq U_p$ under time-varying conditions \cite{Shockley_PhysRev_1952,Blakemore_book_1962}. As a result, an electron (or hole) may be trapped for a certain period before recombining, which tends to unbalance the electron and hole densities. Recent works on nano-waveguides \cite{Aldaya_Optica_2017} and microring resonators \cite{Novarese_OptExpress_2022} have pointed out that modeling the electron and hole recombination rates with separate formulations \cite{Blakemore_book_1962} is necessary to accurately reproduce the nonlinear carrier dynamics observed experimentally. In particular, the carrier decay rate due to trap-assisted recombination generally depends on the initial carrier density \cite{Ahrenkiel_JAP_1991,Aldaya_Optica_2017}, which makes the process nonlinear.

In practice, one may assume $U_n = U_p$ if the dynamics of the trapped electrons and holes is sufficiently slow compared to the
drift-diffusion dynamics, and, in general, to the timescale of interest \cite{Pierret_book_2002,Vasileska_book_2010}. This quasi-stationary approximation is justified when the trap density is small compared to the excess electron and hole densities \cite{Aldaya_Optica_2017}, and typically satisfactory at carrier densities around $10^{15}\sim10^{18}\,\mathrm{cm^{-3}}$ \cite{Borghi_OptExpress_2021}, which we are interested in. Furthermore, the carrier lifetimes due to trap-assisted recombination are typically larger than $1\,\mathrm{ns}$ \cite{Aldaya_Optica_2017,Novarese_OptExpress_2022}. This further corroborates the quasi-stationary approximation on much shorter timescales, where carrier diffusion is expected to dominate. Trap-assisted recombination due to defects at the interface between the semiconductor material and surrounding cladding (so-called surface recombination) \cite{Pierret_book_2002} is discussed later, but similar considerations can be made. 

We consider photonic cavities realized using semiconductor materials that are either intrinsic or homogeneously doped. Therefore, without an external excitation, the electron and hole densities are uniform throughout the cavity. In the presence of an optical excitation pulse, electrons and holes are generated and initially diffuse at different speeds, due to the different mobilities. Electrons and holes, however, are charged particles. As a result, an internal electric field arises, which tends to retard the electrons and accelerate the holes \cite{McKelvey_book_1966}. If trap-assisted recombination does not unbalance the electron and hole densities (an assumption whose limitations have been discussed above), the internal electric field ensures local neutrality. This means that the excess electron density is balanced by an equal excess hole density. Consequently, electrons and holes end up diffusing together, in a so-called ambipolar diffusion regime \cite{VanRoosbroeck_PhysRev_1953}. 

It is clear that if local neutrality were fulfilled exactly, no internal field would be set up. However, the difference between electron and hole densities required to induce the internal field is so small compared to the excess densities themselves, that local neutrality is often a good approximation \cite{McKelvey_book_1966}.

Assuming local neutrality, we denote by $N(\mathbf{r},t)$ the excess density of electrons and holes. We multiply \Eqref{eq:App0:DD-eqs-1} by $p\mu_p$ and \Eqref{eq:App0:DD-eqs-2} by $n\mu_n$, with $\mu_p$ and $\mu_n$ being the hole and electron mobility, respectively. Summing the two equations and using Einstein's relations for the electron and hole diffusion coefficients, we arrive at \cite{McKelvey_book_1966}
\begin{equation}
\label{eq:App0:ambi-eq-DER1}
\frac{\partial N}{\partial t} = D_{\mathrm{eff}}\nabla^2N - \mu_{\mathrm{eff}}\mathbf{E}_0\cdot\nabla N- U + G
\end{equation}
where the effective diffusion coefficient, $D_{\mathrm{eff}}$, is
\begin{equation}
\label{eq:App0:Deff-DER1}
D_{\mathrm{eff}} = \frac{(n_0 + p_0 + 2N)D_nD_p}{(n_0 + N)D_n + (p_0 + N)D_p}
\end{equation}
and the effective mobility, $\mu_{\mathrm{eff}}$, reads 
\begin{equation}
\label{eq:App0:mueff-DER1}
\mu_{\mathrm{eff}} = \frac{\mu_n\mu_p(n_0-p_0)}{(n_0 + N)\mu_n + (p_0 + N)\mu_p}
\end{equation}
Here, $n_0$ ($p_0$) is the electron (hole) density at thermal equilibrium (with $n=n_0+N$ and $p=p_0+N$), whereas $D_n$ and $D_p$ are the electron and hole diffusion coefficient, respectively. The electric field, $\mathbf{E}_0$, accounts for the internal field and any externally applied bias. The external field is assumed to be weak enough that the ambipolar approximation would hold. \Eqref{eq:App0:ambi-eq-DER1} is the basis of the classical Haynes-Shockley experiment to measure the minority carrier mobility in n- or p-type semiconductors \cite{Haynes_PRB_1951,McKelvey_JAP_1956}.

We assume \textbf{high-level injection}, whereby the excess electron and hole density is much larger than the thermal equilibrium concentrations. This is the typical scenario for lasing \cite{Coldren-book2ndEd-2012} and optical switching \cite{Nozaki_NatPhot_2010,Yu_OptExpress_2013,Moille_LPR_2016} applications unless the semiconductor material is heavily doped \cite{Moille_OptLett_2017}. For the sake of reference, we note that the room-temperature equilibrium carrier concentration of intrinsic indium phosphide is around $10^7\,\mathrm{cm^{-3}}$ \cite{Levinshtein-handbook-1998}. We also assume no externally applied bias. Therefore, $\mathbf{E}_0$ reduces to the internal field. For $N\gg n_0,p_0$, \Eqref{eq:App0:mueff-DER1} gives $\mu_{\mathrm{eff}}\approx0$, whereas \Eqref{eq:App0:Deff-DER1} reduces to
\begin{equation}
\label{eq:App0:Deff-high-inj}
D_{\mathrm{eff}} = \frac{2D_nD_p}{D_n+D_p}
\end{equation}

As for the SRH recombination rate, it should be noted that unless the trap level is close enough to the intrinsic Fermi level (around the middle of the electronic band gap), either $n_1$ or $p_1$ will become large, significantly reducing the recombination rate. For the traps to be effective, they must introduce energy levels close to the middle of the gap \cite{Coldren-book2ndEd-2012}, in which case one finds $p_1\approx n_1 \approx n_i$. Thus, \Eqref{eq:App0:SRH-rate} under high-level injection gives \cite{Coldren-book2ndEd-2012}
\begin{equation}
\label{eq:App0:SRH-rate-high-inj}
U \approx N/\tau_{\mathrm{bulk}}
\end{equation}
with $\tau_{\mathrm{bulk}} = \tau^{\mathrm{SRH}}_{n} + \tau^{\mathrm{SRH}}_{p}$ being the recombination lifetime. 
Consequently, \Eqref{eq:App0:ambi-eq-DER1} reduces to the usual form of the ambipolar diffusion equation \cite{McKelvey_book_1966,Yu_OptExpress_2013,Moille_PRA_2016}
\begin{equation}
\label{eq:App0:ambi-eq}
\frac{\partial N(\mathbf{r},t)}{\partial t} = D_{\mathrm{eff}}\nabla^2N(\mathbf{r},t) - \frac{N(\mathbf{r},t)}{\tau_{\mathrm{bulk}}} + G(\mathbf{r},t)
\end{equation}
with the effective diffusion coefficient given by \Eqref{eq:App0:Deff-high-inj}. 

The drift-diffusion equations require suitable boundary conditions. In particular, \textbf{surface recombination} due to interface traps (so-called surface states) should be modeled, in general, via the following boundary conditions \cite{Engl_ProcIEEE_1983,Mui_JAP_1995} 
\begin{subequations}
\begin{align}
& \hat{\mathbf{n}}\cdot\mathbf{J}_n = -q\,U_{n_s}  \label{eq:App0:Rsurf-bound-cond-1}\\
& \hat{\mathbf{n}}\cdot\mathbf{J}_p = q\,U_{p_s} \label{eq:App0:Rsurf-bound-cond-2}
\end{align}
\end{subequations}
Here, $U_{n_s}$ and $U_{p_s}$ are the electron and hole surface recombination rates per unit area, respectively, whereas $\hat{\mathbf{n}}$ is the unit vector normal to the semiconductor surface and pointing outwards.

The physical analogies between trap-assisted bulk recombination and surface recombination lead to similar considerations and mathematical descriptions. The density of surface states, however, is naturally expressed as a density per unit area, because interface defects are spread throughout a plane rather than a volume. As a consequence, $\tau^{\mathrm{SRH}}_{n}$ ($\tau^{\mathrm{SRH}}_{p}$) in \Eqref{eq:App0:SRH-rate} becomes $1/v_{n}$ ($1/v_{p}$), with $v_{n}$ and $v_{p}$ being the electron and hole surface recombination velocity, respectively. These velocities are directly proportional to the electron and hole capture cross-sections, respectively, as well as the density of surface states. If the dynamics of the electrons and holes trapped at the surface states is sufficiently slow, the surface recombination rates for electrons and holes are the same. Then, for a single trap energy level, one finds \cite{Pierret_book_2002}
\begin{equation}
\label{eq:App0:surf-rate}
U_{n_s} = U_{p_s} = U_s = \frac{n_sp_s - n_i^2}{(p_s+p_{1_s})/v_{n} + (n_s+n_{1_s})/v_{p}}
\end{equation}
where $n_s$ and $p_s$ are the electron and hole densities per unit volume at the semiconductor surface, respectively. The expressions of $n_{1_s}$ and $p_{1_s}$ are given by \Eqref{eq:App0:n1} and \Eqref{eq:App0:p1}, respectively, but the intrinsic Fermi level and the intrinsic carrier density are evaluated at the surface. The trap level is that of the surface states.   

It should be noted that surface recombination usually involves trap levels distributed in energy throughout the electronic band gap \cite{Pierret_book_2002}, in contrast to bulk recombination. Therefore, one should in principle integrate \Eqref{eq:App0:surf-rate} over all the trap levels, with possibly energy-dependent surface recombination velocities \cite{Pierret_book_2002}. However, the recombination process is dominated by trap levels close to the middle of the gap, as already noted for bulk recombination. Therefore, one may often assume constant surface recombination velocities, equal to the midgap value \cite{Schenk-book-1998}. Then, under the assumption of high-level injection, \Eqref{eq:App0:surf-rate} reduces to \cite{Coldren-book2ndEd-2012} 
\begin{equation}
\label{eq:App0:surf-rate-high-inj}
U_s \approx v_sN_s
\end{equation}
where $N_s$ is the excess electron and hole density at the semiconductor surface. The surface recombination velocity, $v_s$, is given by $1/v_s = 1/v_n + 1/v_p$. It is strongly dependent on the semiconductor material and fabrication process \cite{Nolte_SolidStateElec_1990,Moille_LPR_2016,Black_NanoLett_2017,Higuera-Rodriguez_NanoLett_2017}.

To obtain the boundary condition for \Eqref{eq:App0:ambi-eq}, we multiply \Eqref{eq:App0:Rsurf-bound-cond-1} and \Eqref{eq:App0:Rsurf-bound-cond-2} by $p\mu_p$ and $n\mu_n$, respectively. We subtract the second equation from the first one and express the electron and hole current densities in terms of the corresponding drift-diffusion contributions. We use the ambipolar approximation (namely, $n = n_0 + N$ and $p = p_0 + N$), as well as $U_{n_s} = U_{p_s} = U_s \approx v_sN_s$, finally arriving at
\begin{equation}
\label{eq:App0:ambipolar-approx-bound-cond}
D_{\mathrm{eff}}\,\hat{\mathbf{n}}\cdot\nabla N(\mathbf{r},t) = -v_sN(\mathbf{r},t)
\end{equation}
Here, the excess carrier density is evaluated at the semiconductor surface.  

It should be pointed out that surface states usually act as donor-like or acceptor-like recombination centers and possess a charge depending on the occupation \cite{Aberle_JAP_1992,Mui_JAP_1995,Donchev_ACSApplEnergMat_2018}. The resulting electric field tends to unbalance the electron and hole densities \cite{Mui_JAP_1995} and should in principle be taken into account by self-consistently coupling the surface recombination rates with Poisson's equation \cite{Girish_IEEETransElecDev_1988,Aberle_JAP_1992,Mui_JAP_1995,McIntosh_JAP_2014}. The unbalance, however, is minor if the density of surface states is small compared to the excess electron and hole densities \cite{Mui_JAP_1995}, and may be neglected, as a first approximation, under high-level injection conditions.

We also note that surface recombination in nanostructures usually dominates over bulk recombination, owing to the large surface-to-volume ratio \cite{Dimitropoulos_APL_2005,Higuera-Rodriguez_NanoLett_2017,Aldaya_Optica_2017}. Therefore, for a given density of surface states, the unbalance between the electron and hole surface recombination rates may eventually become important when the surface-to-volume ratio is sufficiently large \cite{Aldaya_Optica_2017}, and thus jeopardize the ambipolar approximation. This modeling scenario falls outside the scope of this article and is left to future works. 

Here, we model the carrier diffusion by \Eqref{eq:App0:ambi-eq}, and surface recombination by \Eqref{eq:App0:ambipolar-approx-bound-cond}, consistently with previous works \cite{Nozaki_NatPhot_2010,Yu_OptExpress_2013,Moille_LPR_2016}. Electrostatic and saturation effects \cite{Girish_IEEETransElecDev_1988,Aberle_JAP_1992,Mui_JAP_1995} are ignored, as discussed above. Unless otherwise specified, we assume parameters realistic for Indium Phosphide (InP): $D_{\mathrm{eff}} = 10\,\mathrm{cm^2/s}$ \cite{Levinshtein-handbook-1998} and $v_s = 10^4\,\mathrm{cm/s}$ \cite{Yu_OptExpress_2013}. 
  

%% file: sections/section03.tex
\section{Eigenmode expansion and response function formalism} \label{sec: effective carrier density - eig exp}

In optical switches based on micro- or nanoscale semiconductor resonators, the shift in the cavity resonance due to various carrier-induced effects (mainly, free carrier-induced dispersion) \cite{Bennett_JQE_1990} scales to first order with a mode-averaged (or effective) carrier density \cite{Johnson_OE_2006,Yu_OptExpress_2013}:
\begin{equation}
\label{eq:sec3:Neff:def}
N_{\mathrm{eff}}(t) = \frac{ \int_{V_m}N(\mathbf{r},t)\epsilon(\mathbf{r})|\mathbf{E}(\mathbf{r})|^2 d^3\mathbf{r} }{ \int_{V}\epsilon(\mathbf{r})|\mathbf{E}(\mathbf{r})|^2d^3\mathbf{r} } 
\end{equation}
Here, $\mathbf{E}$ is the electric field of the cavity mode, $\epsilon(\mathbf{r})$ is the permittivity and $V_m$ is the semiconductor volume. The integration volume, $V$, encloses the optical cavity with a margin of a few wavelengths. 
The carrier density, $N(\mathbf{r},t)$, follows the ambipolar diffusion equation (\Eqref{eq:App0:ambi-eq}), with surface recombination boundary conditions (\Eqref{eq:App0:ambipolar-approx-bound-cond}).

It should be noted that the field distribution diverges in space at sufficiently large distances from the cavity, due to the leaky nature of the cavity mode \cite{Kristensen_AdvOptPhot_2020}. In principle, this divergence makes the denominator of \Eqref{eq:sec3:Neff:def} ill-defined and raises the question of the normalization of the cavity mode \cite{Kristensen_PRA_2015,Muljarov_PRA_2017,Kristensen_PRA_2017}. The choice of $V$, however, does not affect the decay rate of $N_{\mathrm{eff}}(t)$, which makes the issue unimportant for the scope of this article. We also note that the numerator of \Eqref{eq:sec3:Neff:def} is instead well-defined.   

In the framework of coupled-mode theory, the carrier generation rate in the diffusion equation is a separable function of space and time \cite{Barclay_OptExpress_2005,Moille_PRA_2016}:
\begin{equation}
\label{eq:Sec3:G-separation}
G(\mathbf{r},t) = F(\mathbf{r})f(t)
\end{equation}
We focus on optical switches with carriers generated by two-photon absorption (TPA) \cite{Barclay_OptExpress_2005,Yu_OptExpress_2013,Moille_LPR_2016}. In this case, $f(t)$ describes the time variation of the squared optical intensity inside the cavity. The carrier density excitation profile, $F(\mathbf{r})$, is given by
\begin{equation}
\label{eq:Sec3:F}
F(\mathbf{r}) = |\mathbf{E}(\mathbf{r})|^4
\end{equation}
The case of linear absorption, as relevant for semiconductor lasers \cite{Coldren-book2ndEd-2012}, corresponds to $F(\mathbf{r}) = |\mathbf{E}(\mathbf{r})|^2$. We emphasize that the nonlinear switching dynamics, as analyzed, for instance, in other works \cite{Yu_OptExpress_2013,Moille_LPR_2016,Saldutti_IEEE_2022}, is outside the scope of this article, where we focus on the diffusion of carriers. 
Nonlinear effects on $f(t)$ are neglected, and the carrier diffusion model is linear.  

The carrier density may be expressed as \cite{Moille_PRA_2016}
\begin{equation}
\label{eq:Sec3:forced-evol:N}
N(\mathbf{r},t) = \int_0^t f(\tau) h(\mathbf{r},t-\tau) d\tau  
\end{equation}
where $h(\mathbf{r},t)$ is the response function. The response function (see Sec.$\,$I in the Supplementary Material) is the solution of the homogeneous diffusion equation 
\begin{equation}
\label{eq:Sec3:ambipolar-eq-homog}
\frac{\partial N(\mathbf{r},t)}{\partial t} = D_{\mathrm{eff}}\nabla^2N(\mathbf{r},t) - \frac{N(\mathbf{r},t)}{\tau_{\mathrm{bulk}}} 
\end{equation}
with the initial condition given by $N(\mathbf{r},t=0) = F(\mathbf{r})$. 

To gain insight, we postulate a solution in the form
\begin{equation}
\label{eq:Sec3:ambipolar-eq-homog_post-sol}
N(\mathbf{r},t) = u(\mathbf{r}_{\parallel})v(z)T(t)e^{-\frac{t}{\tau_{\mathrm{bulk}}}}
\end{equation}
with $\mathbf{r}_{\parallel}$ being the in-plane position vector and $z$ the out-of-plane position coordinate. We emphasize that for the solution to be exact, the initial condition should be also separable, namely
\begin{equation}
\label{eq:Sec3:init-cond-separation}
F(\mathbf{r}) = F_{\mathrm{xy}}(\mathbf{r}_{\parallel})F_{\mathrm{z}}(z)
\end{equation}
This is not generally the case. Nonetheless, as we shall see, this separation procedure provides a good approximation of the effective carrier density. 

By inserting \Eqref{eq:Sec3:ambipolar-eq-homog_post-sol} into \Eqref{eq:Sec3:ambipolar-eq-homog}, the latter is reduced to two eigenvalue problems, corresponding to the in-plane and out-of-plane diffusion dynamics, respectively. By expanding the two solutions on the corresponding sets of eigenmodes (see Sec.$\,$II in the Supplementary Material), and averaging over the optical mode profile, one finally arrives at the response function of the \textit{effective} carrier density  
\begin{equation}
\label{eq:Sec3:heff-separation}
h_{\mathrm{eff}}(t) = h_{\mathrm{eff,xy}}(t)h_{\mathrm{eff,z}}(t)e^{-\frac{t}{\tau_{\mathrm{bulk}}}}  
\end{equation}
with the effective carrier density given by
\begin{equation}
\label{eq:Sec3:Neff:forced-evol}
N_{\mathrm{eff}}(t) = \int_0^t f(\tau) h_{\mathrm{eff}}(t-\tau) d\tau  
\end{equation}
Nonlinear effects may be easily included by coupling $N_{\mathrm{eff}}(t)$ with the coupled-mode theory equations describing the optical intensity inside the cavity \cite{Yu_OptExpress_2013,Moille_LPR_2016}. 

\begin{figure}[ht]
\centering\includegraphics[width=1\linewidth]{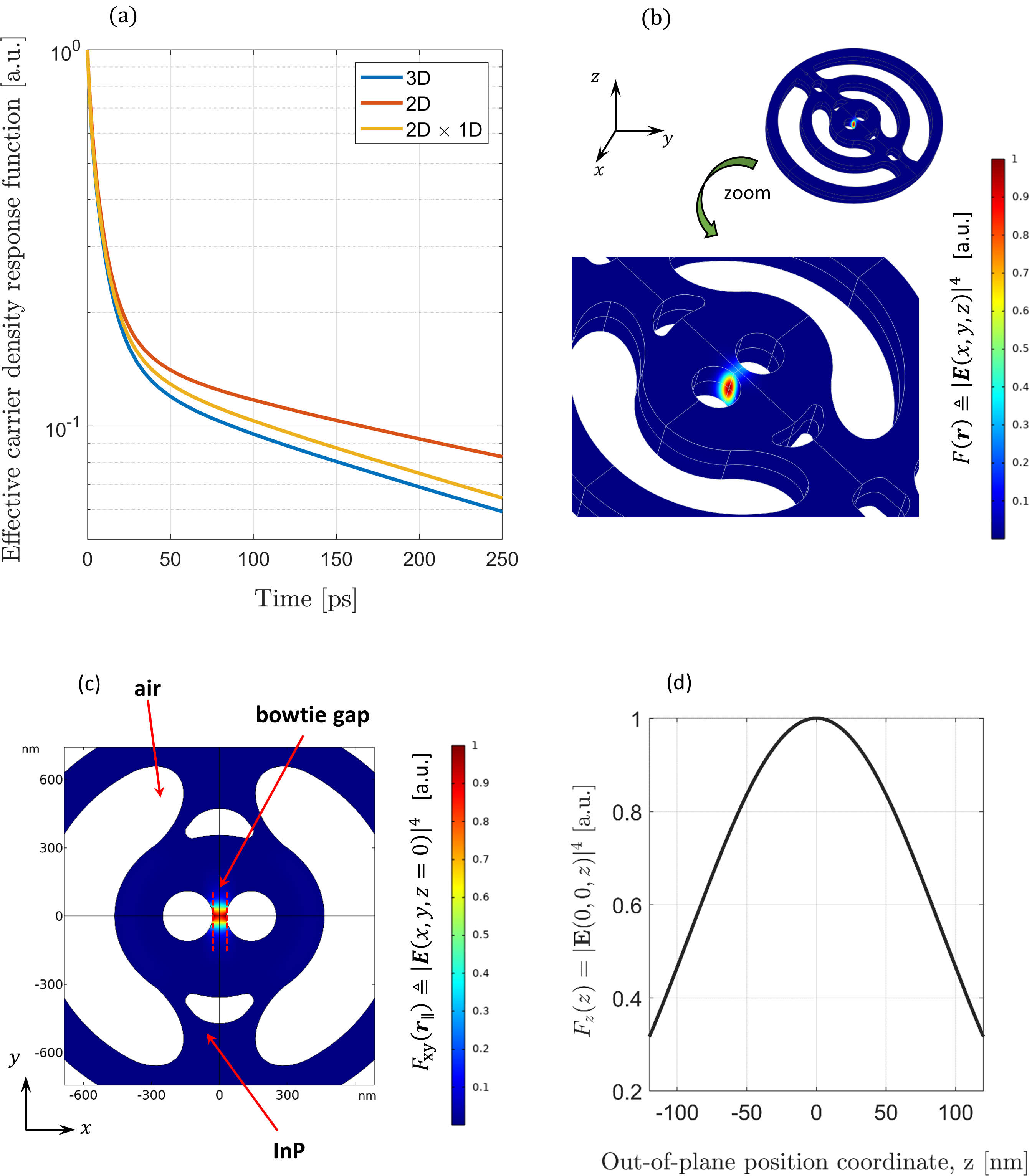}
\caption{\label{fig:sec:SEC3:FIG1} (a) Response function of the effective carrier density for the EDC ring cavity in \tabref{tab:sec:SEC1:TAB1}. The surface recombination velocity is $10^4\,\mathrm{cm/s}$. Results are obtained from 3D simulations (blue), simulations of the in-plane diffusion (red), and combined simulations of the in-plane and out-of-plane diffusion (yellow). The other figures illustrate the carrier density excitation profile for (b) 3D diffusion, (c) in-plane diffusion, and (d) out-of-plane diffusion.}
\end{figure} 
The response function of the in-plane effective carrier density, $h_{\mathrm{eff,xy}}(t)$, reads
\begin{equation}
\label{eq:Sec3:heffxy:Green-func}
h_{\mathrm{eff,xy}}(t) = \sum_i \kappa_{\mathrm{xy}_i}\,e^{-\frac{t}{\tau_{\mathrm{xy}_i}}}
\end{equation}
with $1/\tau_{\mathrm{xy}_i}$ being the eigenvalues of the in-plane diffusion problem, and $\tau_{\mathrm{xy}_i}$ the corresponding lifetimes. The excitation coefficients, $\kappa_{\mathrm{xy}_i}$, are given by
\begin{equation}
\label{eq:Sec3:Neff:kappaxyi}
\kappa_{\mathrm{xy}_i} = A_{\mathrm{xy}_i} \, \frac{ \int_{S_m}u_i(\mathbf{r}_{\parallel})\epsilon(\mathbf{r}_{\parallel})|\mathbf{E}(\mathbf{r}_{\parallel})|^2 d^2\mathbf{r}_{\parallel} }{ \int_{S}\epsilon(\mathbf{r}_{\parallel})|\mathbf{E}(\mathbf{r}_{\parallel})|^2d^2\mathbf{r}_{\parallel} }
\end{equation}
with $A_{\mathrm{xy}_i}$ reading 
\begin{equation}
\label{eq:Sec3:Neff:Axyi}
A_{\mathrm{xy}_i} = \frac{ \int_{S_m} u_i(\mathbf{r}_{\parallel})F_{\mathrm{xy}}(\mathbf{r}_{\parallel})d^2\mathbf{r}_{\parallel}  } { \int_{S_m} u_i^2(\mathbf{r}_{\parallel})d^2\mathbf{r}_{\parallel} }
\end{equation}
and $u_i(\mathbf{r}_{\parallel})$ being the eigenmodes of the in-plane diffusion problem. $S_m$ is the in-plane cross-section area limited to the semiconductor material. $S$ is an in-plane cross-section area enclosing the cavity with a margin of a few wavelengths and including both the semiconductor and surrounding cladding. Similar considerations apply to the denominator of \Eqref{eq:Sec3:Neff:kappaxyi} as to that of \Eqref{eq:sec3:Neff:def}. To arrive at \Eqref{eq:Sec3:Neff:kappaxyi}, we have also assumed the electric energy density in \Eqref{eq:sec3:Neff:def}, $\epsilon(\mathbf{r})|\mathbf{E}(\mathbf{r})|^2$, to be a separable function of $\mathbf{r}_{\parallel}$ and $z$, consistently with the initial condition (cf. \Eqref{eq:Sec3:init-cond-separation}).     

Eigenmodes and eigenvalues are found by solving the diffusion equation in two dimensions via the finite-element method (FEM), and $h_{\mathrm{eff,xy}}(t)$ is readily computed by applying \Eqref{eq:Sec3:heffxy:Green-func}. Alternatively, one may find $h_{\mathrm{eff,xy}}(t)$ by solving the diffusion equation in space and time \cite{Nozaki_NatPhot_2010,Yu_OptExpress_2013,Moille_PRA_2016} and with the initial condition given by $F_{\mathrm{xy}}(\mathbf{r}_{\parallel})$. \Eqref{eq:Sec3:heffxy:Green-func} converges to the result of the latter approach when a sufficiently large number of eigenmodes is considered, as we shall see in \secref{sec: in-plane diffusion}. 

Similar considerations apply to the response function of the out-of-plane effective carrier density, $h_{\mathrm{eff,z}}(t)$. Specifically, one finds   
\begin{equation}
\label{eq:Sec3:heffz:Green-func}
h_{\mathrm{eff,z}}(t) = \sum_j \kappa_{\mathrm{z}_j}\,e^{-\frac{t}{\tau_{\mathrm{z}_j}}}
\end{equation}
with $\tau_{\mathrm{z}_j}$ being the out-of-plane diffusion lifetimes. The corresponding excitation coefficients, $\kappa_{\mathrm{z}_j}$, are found by applying \Eqref{eq:Sec3:Neff:kappaxyi} and \Eqref{eq:Sec3:Neff:Axyi} with obvious changes of the surface integrals into one-dimensional integrals along $z$. In particular, $F_{\mathrm{xy}}(\mathbf{r}_{\parallel})$ and $u_i(\mathbf{r}_{\parallel})$ are replaced by $F_{\mathrm{z}}(z)$ and $v_j(z)$, respectively, the latter being the out-of-plane eigenmodes. We emphasize that \Eqref{eq:Sec3:heffz:Green-func} is a generalization of the usual single-lifetime approximation, $h_{\mathrm{eff,z}}(t) = \exp\left(-t/\tau_{\mathrm{z}_0}\right)$, of the out-of-plane diffusion dynamics, with $1/\tau_{\mathrm{z}_0} = 2v_s/L_z$ \cite{Nozaki_NatPhot_2010,Yu_OptExpress_2013}. Here, $L_z$ is the thickness of the cavity, and $v_s$ is the surface recombination velocity. The out-of-plane diffusion dynamics is analyzed in \secref{sec: out-of-plane diffusion}.       

For example, \figref{fig:sec:SEC3:FIG1}a shows the response function of the effective carrier density for the EDC ring cavity in \tabref{tab:sec:SEC1:TAB1}. The excitation profile, $F(\mathbf{r}) = |\mathbf{E}(\mathbf{r})|^4$, is displayed in \figref{fig:sec:SEC3:FIG1}b. The electric field corresponds to the cavity fundamental mode. It is obtained from three-dimensional FEM simulations of the source-free Maxwell's equations in the frequency domain with radiation boundary conditions \cite{Kountouris-OptExress-2022}. The excitation profile over the in-plane cross-section area ($z=0$) and along the cavity growth direction, $z$, at the center of the cavity ($x=y=0$) is shown in \figref{fig:sec:SEC3:FIG1}c and \figref{fig:sec:SEC3:FIG1}d, respectively. To focus on the effect of carrier diffusion, bulk recombination is ignored ($\tau_{\mathrm{bulk}}=\infty$). The curves in \figref{fig:sec:SEC3:FIG1}a correspond to full three-dimensional simulations (blue), two-dimensional simulations of the in-plane diffusion (red, \Eqref{eq:Sec3:heffxy:Green-func}), and combined simulations of the in-plane and out-of-plane diffusion (yellow, \Eqref{eq:Sec3:heff-separation}). 
In three dimensions, the response function is found from space- and time-domain simulations, but the eigenmode approach could be also applied (see Sec.$\,$I in the Supplementary Material). 

From \figref{fig:sec:SEC3:FIG1}a, the separation into in-plane and out-of-plane diffusion (2D$\times$1D, yellow) is seen to be a good approximation of the full three-dimensional dynamics (3D, blue). The out-of-plane diffusion mainly affects the long-timescale decay rate. The separation of the excitation profile, \Eqref{eq:Sec3:init-cond-separation}, is fulfilled only approximately, which explains the non-perfect agreement between \Eqref{eq:Sec3:heff-separation} and the three-dimensional simulation results. 

For the other photonic cavities, we have found similar results. We note that the fundamental mode profile of PhC H0 cavities features a node at the center of the cavity (see \figref{fig:sec:SEC1:FIG1-FIG2}c). However, the separation procedure herein illustrated is still applicable by choosing the in-plane position coordinates, $x$ and $y$, around a maximum of the field profile when considering $F_{\mathrm{z}}(z)$.                         

%% file: sections/section04.tex
\section{In-plane diffusion} \label{sec: in-plane diffusion}
 
In the following, we focus on the contribution to the carrier dynamics due to in-plane diffusion. 

\begin{figure}[ht]
\centering\includegraphics[width=1\linewidth]{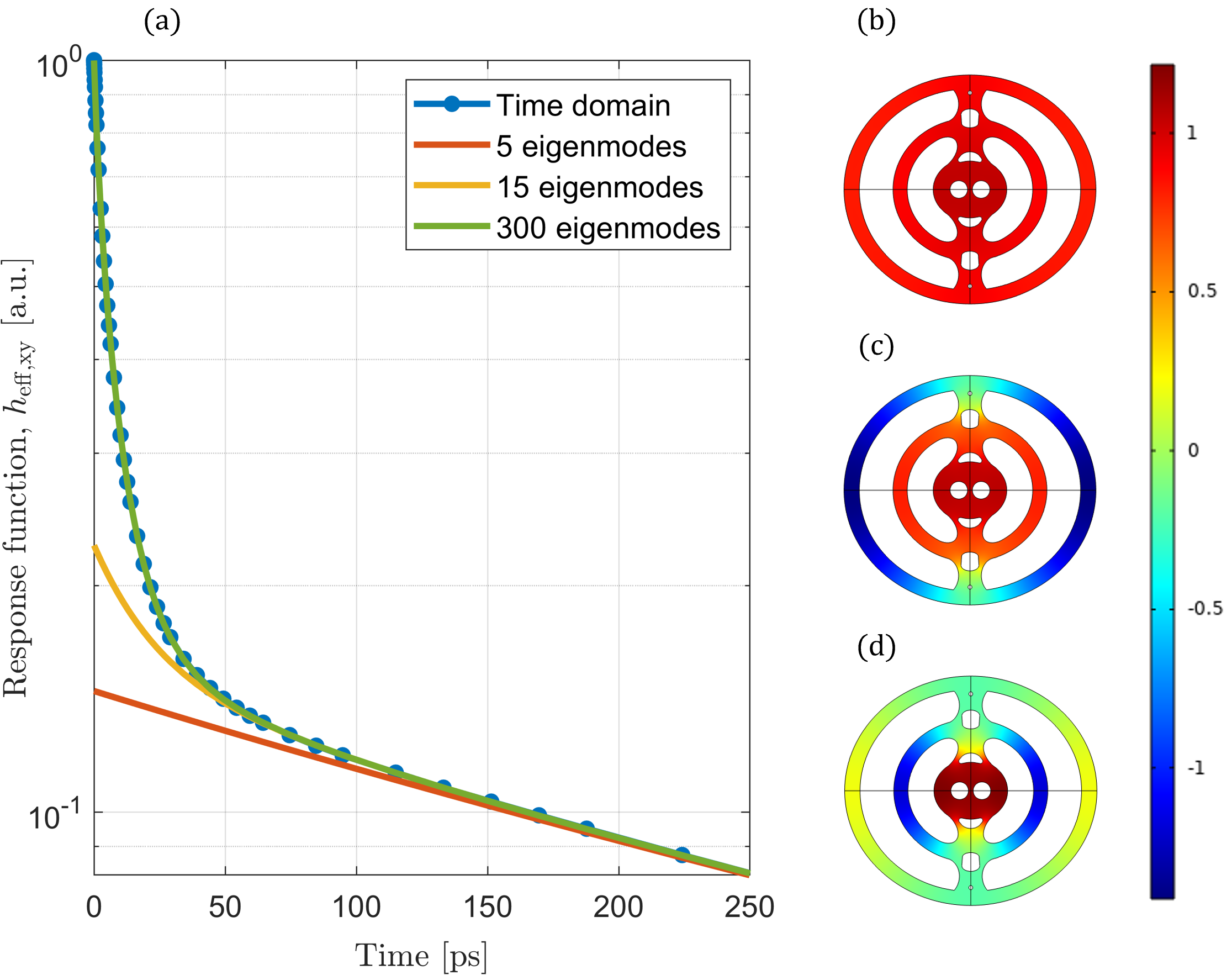}
\caption{\label{fig:sec:SEC4:FIG1} (a) Response function of the in-plane effective carrier density for the EDC ring cavity in \tabref{tab:sec:SEC1:TAB1}. The response function is obtained by space- and time-domain simulations (blue) and applying the eigenmode approach, with the first 5 (red), 15 (yellow), and 300 eigenmodes (green). The eigenmodes are ordered by decreasing values of the diffusion lifetimes. (b) First, (c) second and (d) third eigenmode. The surface recombination velocity is $10^4\,\mathrm{cm/s}$.}
\end{figure}
\figref{fig:sec:SEC4:FIG1}a shows the response function of the in-plane effective carrier density for the EDC ring cavity in \tabref{tab:sec:SEC1:TAB1}, as obtained from space- and time-domain simulations (blue) and applying the eigenmode expansion introduced in \secref{sec: effective carrier density - eig exp}. Here, we have sorted the eigenmodes by decreasing values of the diffusion lifetimes and accounted for the first 5 (red), 15 (yellow), and 300 eigenmodes (green). For example, the first three eigenmodes are shown in \figref{fig:sec:SEC4:FIG1}b, c and d, respectively. The eigenmode approach conveniently captures the long-timescale decay rate with only a few eigenmodes, which is advantageous compared to time-domain simulations. 
Empirically, we have found that around $300$ eigenmodes are generally enough to calculate the response function at all times. This criterion applies to all the photonic cavities considered in this article. Moreover, as we shall see, sorting the eigenmodes by different criteria may reduce the number of eigenmodes being required. 

\begin{figure}[ht]
\centering\includegraphics[width=1\linewidth]{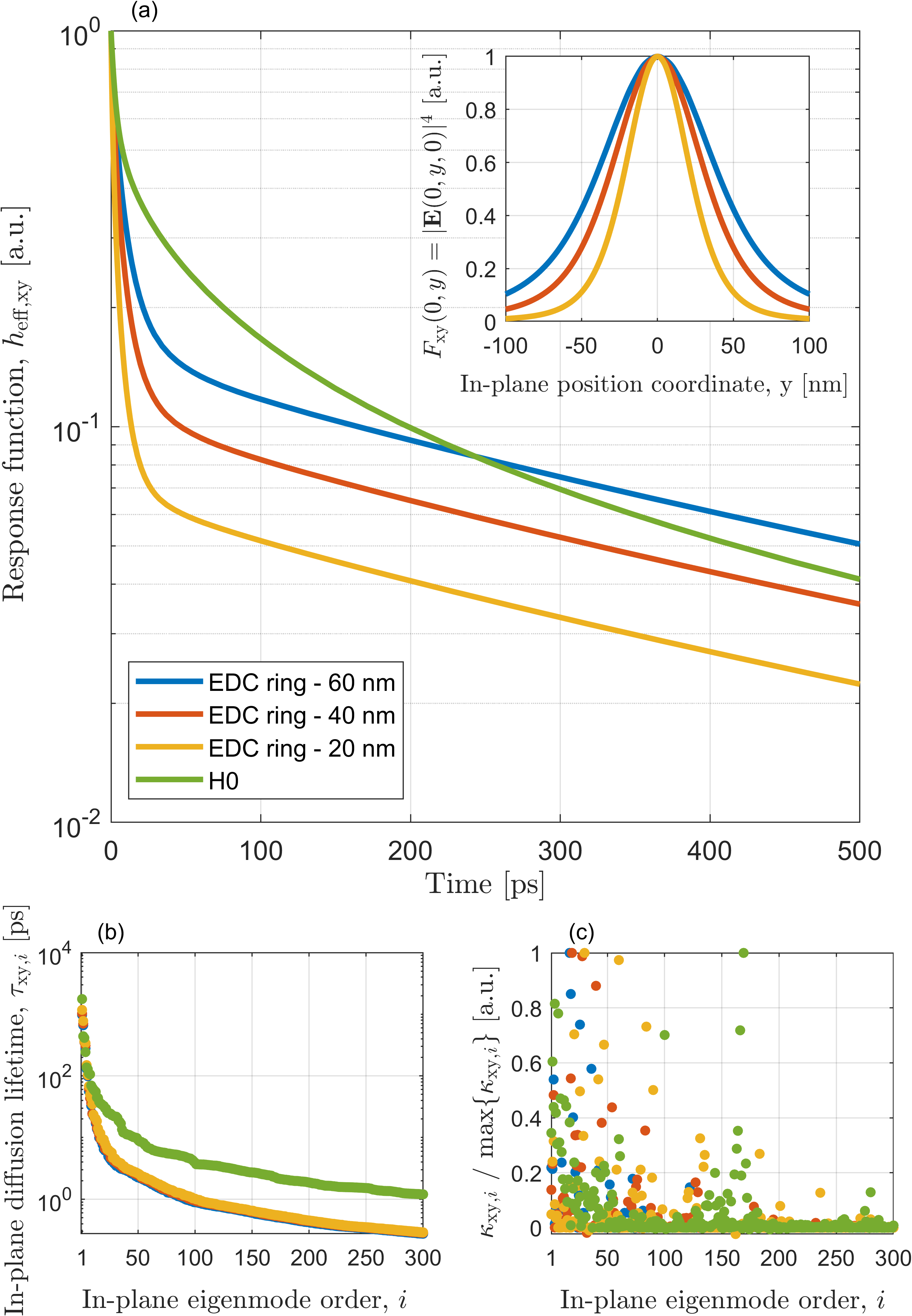}
\caption{\label{fig:sec:SEC4:FIG2} (a) Response function of the in-plane effective carrier density for the PhC H0 cavity (green) and EDC ring cavity in \figref{fig:sec:SEC1:FIG1-FIG2}. The bowtie gap is $60\,\mathrm{nm}$ (blue), $40\,\mathrm{nm}$ (red) and $20\,\mathrm{nm}$ (yellow). The surface recombination velocity is $10^4\,\mathrm{cm/s}$. Inset: carrier density excitation profile along the y-direction in the EDC ring cavities. (b) In-plane diffusion lifetimes and (c) excitation coefficients ordered by decreasing values of the diffusion lifetimes. For each cavity, the excitation coefficients are normalized to the maximum value.}
\end{figure} 
As shown in \figref{fig:sec:SEC1:FIG1-FIG2}a, tighter field confinement accelerates the decay rate of the effective carrier density. Therefore, exploring the impact of the bowtie gap in EDC cavities is interesting. For this purpose, \figref{fig:sec:SEC4:FIG2}a displays the response function of the in-plane effective carrier density for the PhC H0 cavity (green) and EDC ring cavity in \figref{fig:sec:SEC1:FIG1-FIG2}. The bowtie gap is $60\,\mathrm{nm}$ (blue), $40\,\mathrm{nm}$ (red) and $20\,\mathrm{nm}$ (yellow). We note that even smaller values have been demonstrated both in Silicon \cite{Albrechtsen_NatComm_2022} and InP \cite{Xiong_CLEO_2023}. Scaling down the bowtie gap accelerates the decay rate on the short timescale. We emphasize that a smaller bowtie gap does not only squeeze the field along the $x$-direction (for the reference system, see \figref{fig:sec:SEC1:FIG1-FIG2}b), but also along the $y$-direction, as illustrated in the inset of \figref{fig:sec:SEC4:FIG2}a. 

Unfortunately, direct inspection of the diffusion lifetimes, $\tau_{\mathrm{xy}_i}$ (\figref{fig:sec:SEC4:FIG2}b), or the excitation coefficients, $\kappa_{\mathrm{xy}_i}$ (\figref{fig:sec:SEC4:FIG2}c), does not elucidate further the role of the bowtie gap. Furthermore, the size of the bowtie gap barely affects the TPA and FCA mode volumes (not shown). This observation signifies that these mode volumes are not a direct measure of the carrier diffusion enhancement.     

To gain insight, we introduce the in-plane instantaneous lifetime, $\tau_{\mathrm{eff,xy}}(t)$, defined as follows:
\begin{equation}
\label{eq:Sec4:taueffxy:def}
\frac{1}{\tau_{\mathrm{eff,xy}}(t)} = -\frac{d\ln\left[h_{\mathrm{eff,xy}}(t)\right]}{dt}
\end{equation}
In particular, by inserting the eigenmode expansion from \Eqref{eq:Sec3:heffxy:Green-func}, one finds 
\begin{equation}
\label{eq:Sec4:taueffxy:eig}
\frac{1}{\tau_{\mathrm{eff,xy}}(t)} = \frac{\sum_i \frac{\kappa_{\mathrm{xy}_i}}{\tau_{\mathrm{xy}_i}}\,e^{-\frac{t}{\tau_{\mathrm{xy}_i}}}}{\sum_i \kappa_{\mathrm{xy}_i}\,e^{-\frac{t}{\tau_{\mathrm{xy}_i}}}}
\end{equation}
For $t\gg\tau_{\mathrm{xy}_i}$, the instantaneous lifetime reduces to the longest lifetime, $\tau_{\mathrm{xy}_1}$, which supports the definition. At a generic time, the inverse of the instantaneous lifetime gives the exponential decay rate of the effective carrier density.

\begin{figure}[ht]
\centering\includegraphics[width=1\linewidth]{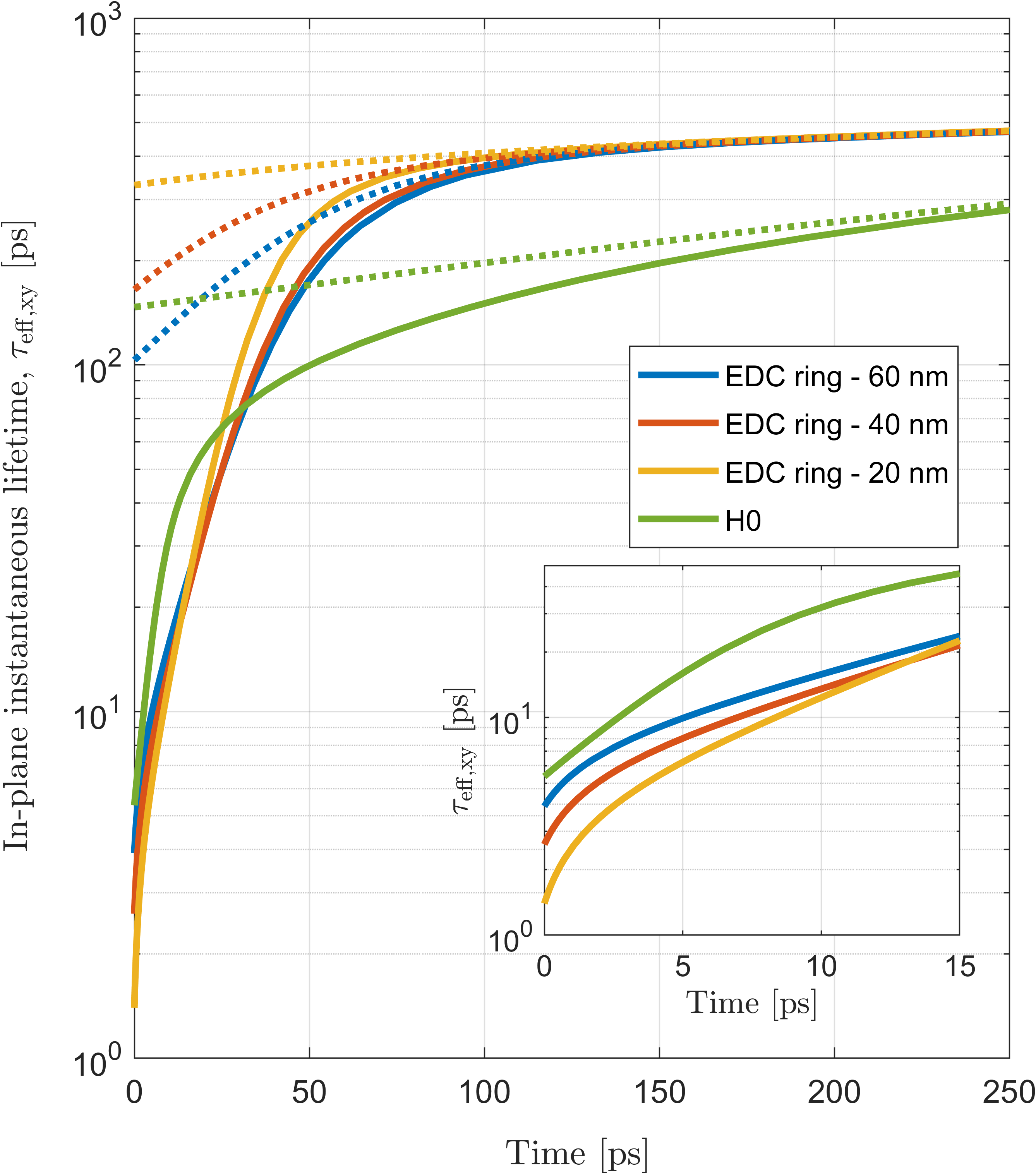}
\caption{\label{fig:sec:SEC4:FIG3} In-plane instantaneous lifetime versus time (solid) for the PhC H0 cavity (green) and EDC ring cavity in \figref{fig:sec:SEC1:FIG1-FIG2}. The bowtie gap is $60\,\mathrm{nm}$ (blue), $40\,\mathrm{nm}$ (red) and $20\,\mathrm{nm}$ (yellow). The dotted lines only account for the eigenmodes with the 10 longest diffusion lifetimes. The surface recombination velocity is $10^4\,\mathrm{cm/s}$. Inset: zoom on the short timescale.}
\end{figure}
\figref{fig:sec:SEC4:FIG3} shows the instantaneous lifetime versus time for the PhC H0 cavity (green) and EDC ring cavity. Again, the bowtie gap is $60\,\mathrm{nm}$ (blue), $40\,\mathrm{nm}$ (red) and $20\,\mathrm{nm}$ (yellow). Including only a few eigenmodes with the longest diffusion lifetime (dotted) is enough to calculate the instantaneous lifetime on the long timescale, as expected. We also note that the diffusion lifetimes decrease with increasing surface recombination velocity (not shown). Consequently, for larger surface recombination velocities even fewer eigenmodes are enough to capture the long-timescale decay rate. Zooming on the short timescale (inset) highlights that a smaller bowtie gap reduces the instantaneous lifetime, which is consistent with \figref{fig:sec:SEC4:FIG2}a. Furthermore, compared to the PhC H0 cavity, the EDC cavities feature shorter instantaneous lifetimes on the short timescale due to the tighter field confinement. The H0 cavity, on the other hand, is faster on the long timescale, most likely due to the larger area exposed to surface recombination.           

It is instructive to consider the instantaneous lifetime at zero time:
\begin{equation}
\label{eq:Sec4:taueffxy:zero-time}
\tau_{\mathrm{eff,xy}}^{-1}(t=0) = \frac{\sum_i \frac{\kappa_{\mathrm{xy}_i}}{\tau_{\mathrm{xy}_i}}}{\sum_i \kappa_{\mathrm{xy}_i}}
\end{equation}
From here, it is evident that not the lifetimes themselves, but instead a \textit{weighted average} of the (inverse) lifetimes determines the instantaneous lifetime, with weights given by the excitation coefficients. This explains why considering either the lifetimes or the excitation coefficients alone (see \figref{fig:sec:SEC4:FIG2}b and \figref{fig:sec:SEC4:FIG2}c) does not emphasize the impact of the geometry on the carrier diffusion dynamics. Furthermore, \Eqref{eq:Sec4:taueffxy:zero-time} suggests that sorting the eigenmodes by their excitation coefficients may be advantageous on the short timescale. 

\begin{figure}[ht]
\centering\includegraphics[width=1\linewidth]{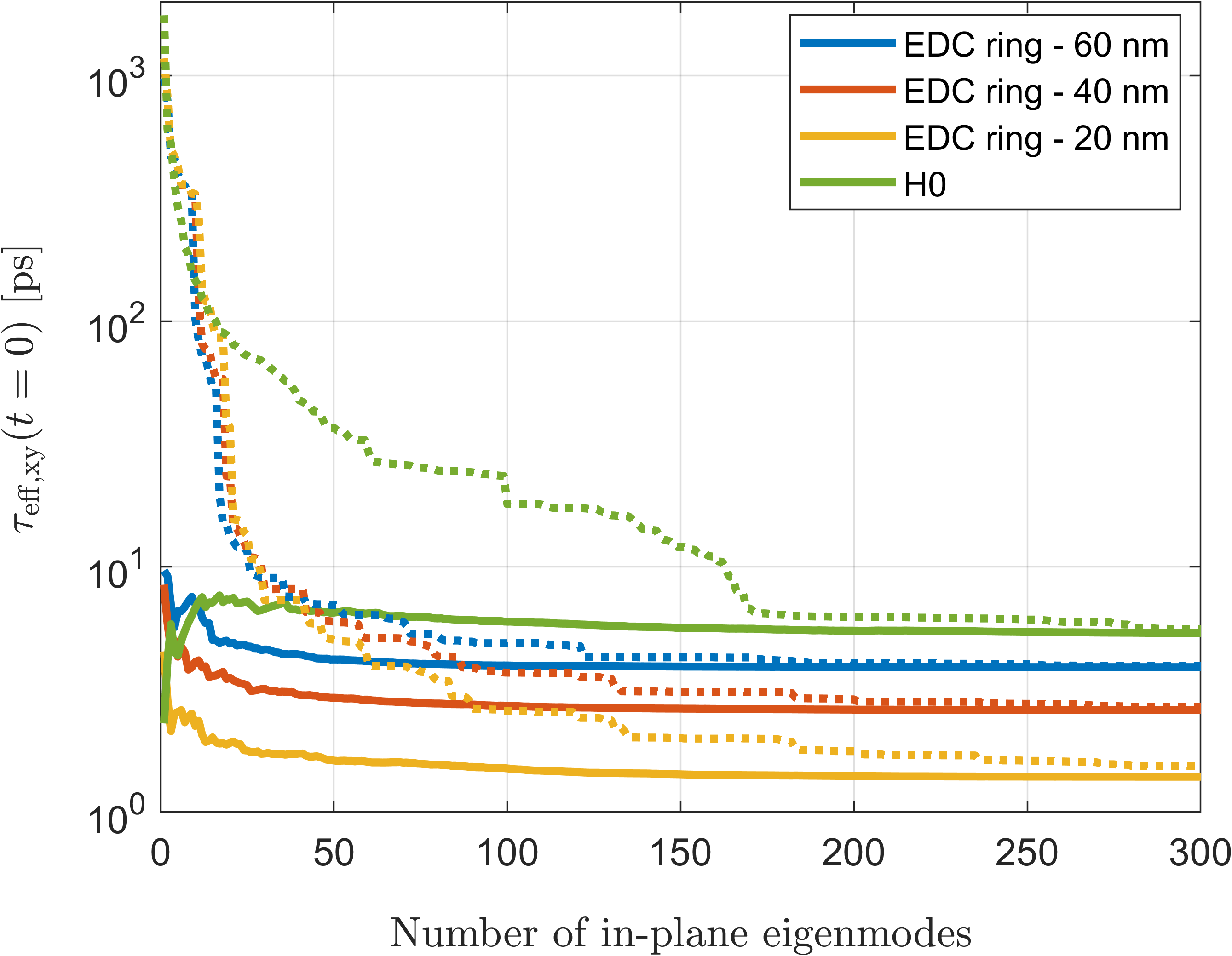}
\caption{\label{fig:sec:SEC4:FIG4} In-plane instantaneous lifetime at zero time versus the number of eigenmodes taken into account. The PhC H0 cavity (green) and EDC ring cavity in \figref{fig:sec:SEC1:FIG1-FIG2} are considered. The bowtie gap is $60\,\mathrm{nm}$ (blue), $40\,\mathrm{nm}$ (red) and $20\,\mathrm{nm}$ (yellow). The eigenmodes are ordered by decreasing absolute values of the excitation coefficients (solid) or decreasing diffusion lifetimes (dotted). The surface recombination velocity is $10^4\,\mathrm{cm/s}$.}
\end{figure}
This is evident from \figref{fig:sec:SEC4:FIG4}, showing the instantaneous lifetime at zero time versus the number of eigenmodes taken into account. Sorting the eigenmodes by decreasing values of $|\kappa_{\mathrm{xy}_i}|$ (solid) reduces the number of eigenmodes to be included for an accurate estimate. In contrast, sorting the eigenmodes by decreasing diffusion lifetimes (dotted) significantly slows down the convergence. We note that sorting the eigenmodes by \textit{increasing} diffusion lifetimes (not shown) is not advantageous either. In fact, the diffusion lifetimes monotonically decrease with increasing order of the eigenmodes (see \figref{fig:sec:SEC4:FIG2}b). Therefore, high-order eigenmodes may easily have diffusion lifetimes much smaller than the instantaneous lifetime at zero time.      

To further quantify the impact of the in-plane geometry, we consider the response of the effective carrier density in the presence of in-plane diffusion only:
\begin{equation}
\label{eq:Sec4:Neffxy:forced-evol}
N_{\mathrm{eff,xy}}(t) = \int_0^t f(\tau) h_{\mathrm{eff,xy}}(t-\tau) d\tau  
\end{equation}
In particular, we assume the time variation, $f(t)$, of the squared optical intensity inside the cavity to be Gaussian 
\begin{equation}
\label{eq:Sec4:Gaussian-pulse}
f(t) = \exp\left[-\frac{(t-t_0)^2}{2\sigma^2}\right] 
\end{equation}
with full width at half maximum (FWHM) given by 
\begin{equation}
\label{eq:Sec4:Gaussian-pulse-duration}
\Delta = 2\sigma\sqrt{2\ln(2)}
\end{equation}
In this case, the in-plane effective carrier density, $N_{\mathrm{eff,xy}}(t)$, can be computed analytically:
\begin{equation}
\label{eq:Sec4:Neffxy:forced-evol-Gaussian-pulse}
N_{\mathrm{eff,xy}}(t) = \frac{\sigma\sqrt{2\pi}}{2}\sum_iH_i(t)\Theta_i(t)
\end{equation}
with the two time-domain factors, $H_i(t)$ and $\Theta_i(t)$, given by 
\begin{subequations}
\begin{align}
H_i(t) & = \exp\left[-\frac{t-t_0-\frac{\sigma^2}{2\tau_{\mathrm{xy,i}}}}{\tau_{\mathrm{xy,i}}}\right]\\
\Theta_i(t) & = \mathrm{erf}\left[\frac{t_0+\frac{\sigma^2}{\tau_{\mathrm{xy,i}}}}{\sigma\sqrt{2}}\right] - \mathrm{erf}\left[\frac{t_0-t+\frac{\sigma^2}{\tau_{\mathrm{xy,i}}}}{\sigma\sqrt{2}}\right]
\end{align}
\end{subequations}
Here, $\mathrm{erf}$ is the error function, $\mathrm{erf}(x) = \frac{2}{\sqrt{\pi}}\int_0^x e^{-y^2}dy$. We note that the intracavity excitation pulse, $f(t)$, only matches the external input pulse if the duration of the latter is much longer than the cavity photon lifetime, as determined by the loaded Q-factor.   

\begin{figure}[ht]
\centering\includegraphics[width=1\linewidth]{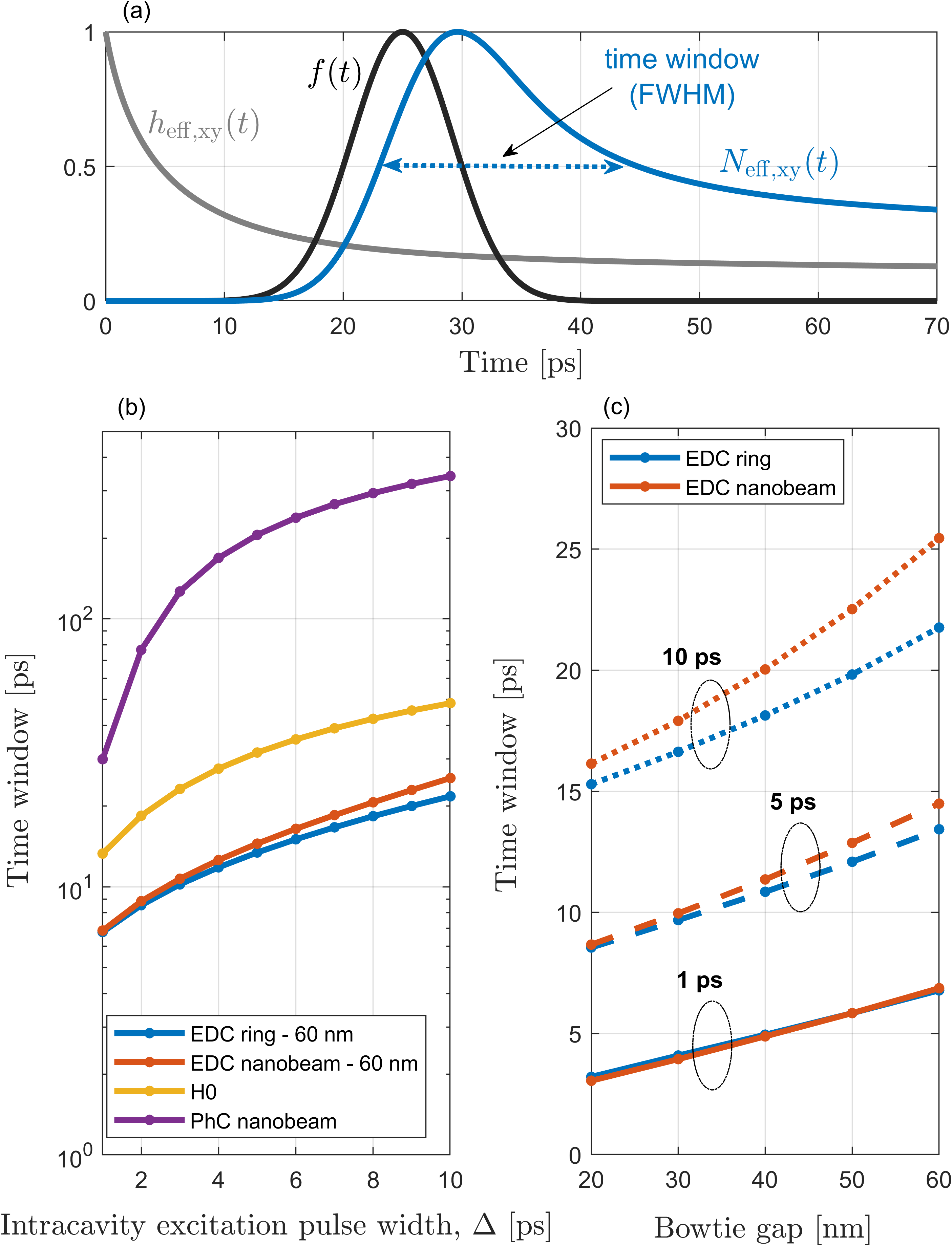}
\caption{\label{fig:sec:SEC4:FIG5} (a) Response function (grey) of the in-plane effective carrier density (blue) in the EDC ring cavity in \tabref{tab:sec:SEC1:TAB1}. The squared optical intensity in the cavity (intracavity excitation) is a Gaussian pulse (black) with full width at half maximum (FWHM) of $10\,\mathrm{ps}$. (b) Time window (FWHM) of the in-plane effective carrier density (cf. \figref{fig:sec:SEC4:FIG5}a) versus the width of the intracavity excitation pulse. The cavities in \tabref{tab:sec:SEC1:TAB1} are considered, with the corresponding colors indicated in the legend. (c) Time window of the in-plane effective carrier density in the EDC ring cavity (blue) and EDC nanobeam cavity (red) versus the size of the bowtie gap. The intracavity excitation pulse width is $1\,\mathrm{ps}$ (solid), $5\,\mathrm{ps}$ (dashed) and $10\,\mathrm{ps}$ (dotted). The surface recombination velocity is $10^4\,\mathrm{cm/s}$.}
\end{figure}
As an example, \figref{fig:sec:SEC4:FIG5}a shows the time evolution of the in-plane effective carrier density (blue) in response to an intracavity Gaussian excitation (black) with FWHM of $10\,\mathrm{ps}$. The EDC ring cavity in \tabref{tab:sec:SEC1:TAB1} is considered, and the corresponding response function (grey) is also shown. The time window (FWHM) of the effective carrier density (cf. \figref{fig:sec:SEC4:FIG5}a) is reported in \figref{fig:sec:SEC4:FIG5}b as a function of the intracavity excitation pulse width. On this timescale, the impact of out-of-plane diffusion is safely negligible, as shown in \secref{sec: out-of-plane diffusion}. We note that the switching time (FWHM of the cavity transmission) \cite{Nozaki_NatPhot_2010,Yu_OptExpress_2013,Ono_NatPhot_2020,Guo_NatPhot_2022} is a critical figure of merit in optical switching applications, and is also influenced by nonlinear effects \cite{Yu_APL_2014}, herein neglected. Yet, a faster linear response generally reduces the switching time. Therefore, \figref{fig:sec:SEC4:FIG5}b suggests that EDC cavities (blue and red) may offer shorter switching times than H0 (yellow) or nanobeam (purple) PhC cavities. Despite the differences in the geometry surrounding the bowtie, the two types of EDC cavities show comparable responses due to the similar field hot spots (see \figref{fig:sec:SEC1:FIG1-FIG2}b and \figref{fig:sec:SEC1:FIG1-FIG2}d). 

In addition, shrinking the bowtie gap may further reduce the switching time, as suggested by \figref{fig:sec:SEC4:FIG5}c. Here, the time window of the effective carrier density in EDC ring cavities (blue) and EDC nanobeam cavities (red) decreases with the bowtie gap. 
As a reference, we note that switching times measured for optical switches based on PhC H0 cavities are on the order of $20$-$50\,\mathrm{ps}$ \cite{Nozaki_NatPhot_2010,Yu_OptExpress_2013}.          

So far, we have assumed a surface recombination velocity of $10^4\,\mathrm{cm/s}$, representative of InP \cite{Yu_OptExpress_2013}. However, this parameter is strongly influenced by the semiconductor material and fabrication process. For instance, the surface recombination velocity may be intentionally increased by growing surface quantum wells on InP photonic cavities \cite{Bazin_APL_2014} or by ion implantation in silicon cavities \cite{Tanabe_APL_2007}. Compared to InP, other materials, such as Gallium Arsenide \cite{Moille_LPR_2016}, may feature much higher surface recombination velocities. On the other hand, surface passivation techniques have proved effective in limiting surface recombination \cite{Black_NanoLett_2017,Higuera-Rodriguez_NanoLett_2017}.   

\begin{figure}[ht]
\centering\includegraphics[width=1\linewidth]{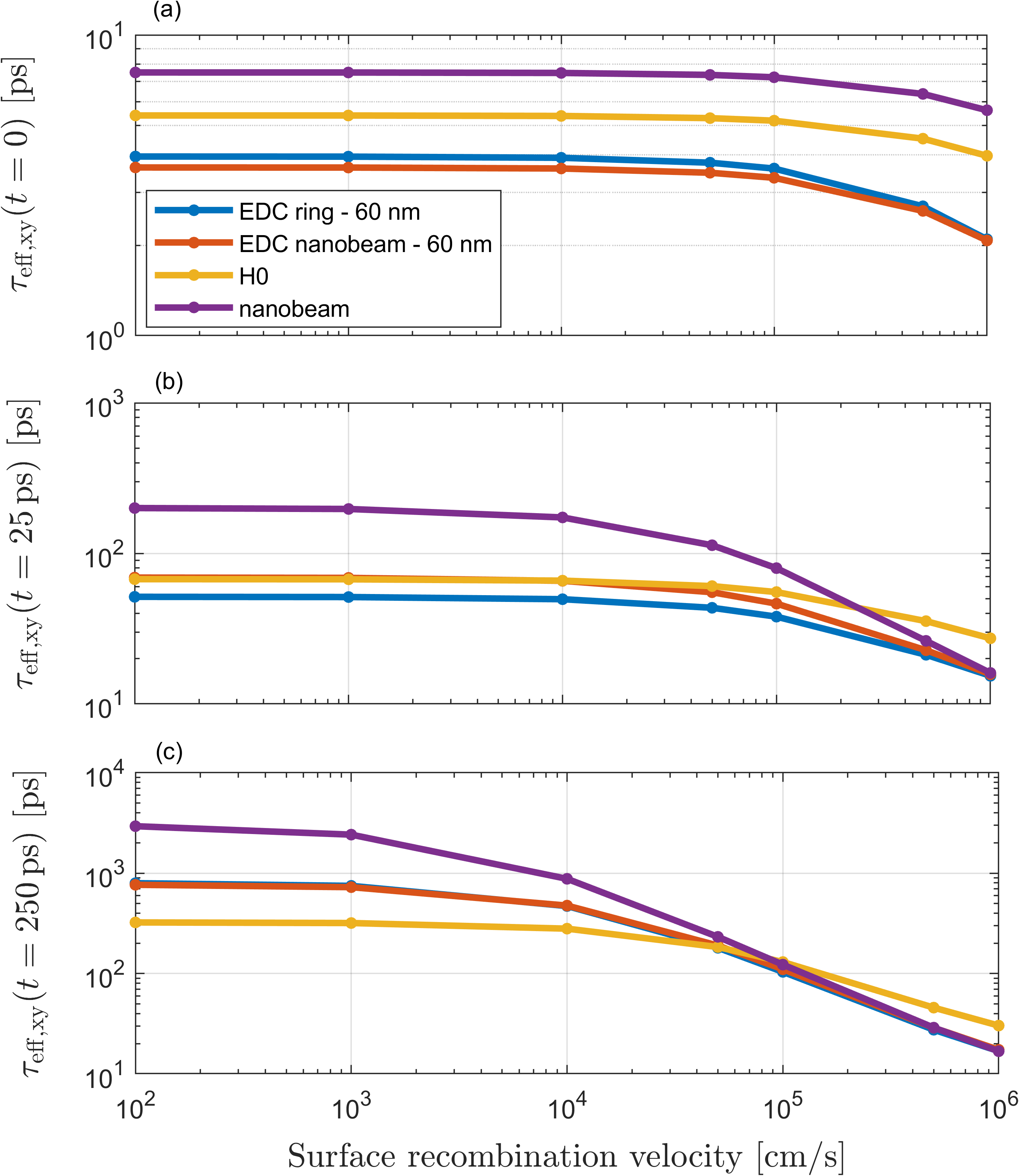}
\caption{\label{fig:sec:SEC4:FIG6} In-plane instantaneous lifetime at (a) zero time, (b) $25\,\mathrm{ps}$ and (c) $250\,\mathrm{ps}$ as a function of the surface recombination velocity. The same cavities as in \figref{fig:sec:SEC4:FIG5}b are considered.}
\end{figure}
\figref{fig:sec:SEC4:FIG6} explores the impact of surface recombination by showing the in-plane instantaneous lifetime at (a) zero time, (b) $25\,\mathrm{ps}$ and (c) $250\,\mathrm{ps}$ as a function of the surface recombination velocity. The different colors correspond to the same cavities as in \figref{fig:sec:SEC4:FIG5}b. For all cavities, the instantaneous lifetime at zero time hardly depends on the surface recombination velocity. This signifies that the carrier diffusion speed is dominated, in the initial stage, by the excitation spatial profile. A stronger dependence on surface recombination is observed with increasing time as diffusion gradually smears out the initial spatial distribution of the carrier density. On the short timescale (\figref{fig:sec:SEC4:FIG6}a and \figref{fig:sec:SEC4:FIG6}b), EDC cavities are generally faster than conventional geometries. Consistently with \figref{fig:sec:SEC4:FIG3}, PhC H0 cavities may offer higher decay rates at longer times (\figref{fig:sec:SEC4:FIG6}c), but the advantage fades with increasing surface recombination. Interestingly, we also note that sufficiently large values of surface recombination lead, at sufficiently long times (\figref{fig:sec:SEC4:FIG6}b and \figref{fig:sec:SEC4:FIG6}c), to the same carrier decay rate in the EDC cavities and PhC nanobeam cavities. This is because the carrier diffusion essentially reduces to a one-dimensional phenomenon, with similar diffusion lengths in the EDC and PhC nanobeam cavities herein considered. The effect is further discussed in \secref{sec: out-of-plane diffusion} in connection with \figref{fig:sec:SEC5:FIG3}.  

We emphasize, though, that large values of surface recombination velocity are not necessarily advantageous for optical switching applications \cite{Moille_LPR_2016}. Indeed, if the carrier diffusion dynamics is much faster than the excitation pulse, the carriers decay before the optical intensity in the cavity has built up significantly. As a result, the nonlinear change in the cavity transmission is weak, and the switching contrast degrades. For a given input power, the switching contrast is optimal when the duration of the excitation pulse matches the short-timescale carrier dynamics \cite{Saldutti_IEEE_2022}.

%% file: sections/section05.tex
\section{Out-of-plane diffusion} 
\label{sec: out-of-plane diffusion}

In the following, we analyze the contribution to the carrier dynamics due to out-of-plane diffusion. In contrast to the case of in-plane diffusion, the out-of-plane diffusion lifetimes and eigenmodes are described by simple analytical formulas thanks to the one-dimensional nature of the problem. 

The out-of-plane diffusion lifetimes, $\tau_{\mathrm{z}_j}$, are given by
\begin{equation}
\label{eq:Sec5:tauz}
\tau_{\mathrm{z}_j} = \frac{1}{D_{\mathrm{eff}}\alpha_{\mathrm{z}_j}^2}
\end{equation}
where $\alpha_{\mathrm{z}_j}$ obeys the following equation (see Sec.$\,$II in the Supplementary Material):
\begin{equation}
\label{eq:Sec5:kz}
\cot{\left(\frac{\alpha_{\mathrm{z}_j}L_{\mathrm{z}}}{2}\right)} = \left(\frac{2D_{\mathrm{eff}}}{L_{\mathrm{z}}v_{\mathrm{s}}}\right)\left(\frac{\alpha_{\mathrm{z}_j}L_{\mathrm{z}}}{2}\right)
\end{equation}
Here, $L_{\mathrm{z}}$ is the cavity thickness along the cavity growth direction, and the surface recombination velocity, $v_s$, is the same at the top and bottom surface of the cavity. 

The diffusion lifetimes and the excitation coefficients, $\kappa_{\mathrm{z}_j}$, determine the response function of the out-of-plane effective carrier density, $h_{\mathrm{eff,z}}(t)$ (cf. \Eqref{eq:Sec3:heffz:Green-func}). The excitation coefficients are given by
\begin{equation}
\label{eq:Sec5:Neff:kappazj}
\kappa_{\mathrm{z}_j} = A_{\mathrm{z}_j} \, \frac{ \int_{L_{\mathrm{z}}}v_j(z)\epsilon(z)|\mathbf{E}(z)|^2 dz }{ \int_{L}\epsilon(z)|\mathbf{E}(z)|^2dz }
\end{equation}
with $v_j(z)$ being the out-of-plane eigenmodes, and $A_{\mathrm{z}_j}$ reading
\begin{equation}
\label{eq:Sec5:Az-cos}
A_{\mathrm{z}_j} = \frac{2\alpha_{\mathrm{z}_j}}{\alpha_{\mathrm{z}_j}L_{\mathrm{z}} + \sin{\left(\alpha_{\mathrm{z}_j}L_{\mathrm{z}}\right)}}\int_{L_{\mathrm{z}}} \cos{\left(\alpha_{\mathrm{z}_j}z\right)}F_{\mathrm{z}}(z)dz  
\end{equation}
The integration line, $L$, in \Eqref{eq:Sec5:Neff:kappazj} includes the whole thickness of the cavity, with a margin of a few wavelengths on the top and bottom. Again, similar considerations apply to the denominator of \Eqref{eq:Sec5:Neff:kappazj} as to that of \Eqref{eq:sec3:Neff:def}. 

In analogy to \Eqref{eq:Sec4:taueffxy:def}, we introduce the out-of-plane instantaneous lifetime, $\tau_{\mathrm{eff,z}}(t) = -d\ln\left[h_{\mathrm{eff,z}}(t)\right]/dt$. From the eigenmode expansion, we obtain    
\begin{equation}
\label{eq:Sec5:taueffz:eig}
\frac{1}{\tau_{\mathrm{eff,z}}(t)} = \frac{\sum_j \frac{\kappa_{\mathrm{z}_j}}{\tau_{\mathrm{z}_j}}\,e^{-\frac{t}{\tau_{\mathrm{z}_j}}}}{\sum_j \kappa_{\mathrm{z}_j}\,e^{-\frac{t}{\tau_{\mathrm{z}_j}}}}
\end{equation}
We emphasize that \Eqref{eq:Sec5:kz} and \Eqref{eq:Sec5:Az-cos} should be generalized if the cavity geometry is not $z$-symmetric or the surface recombination velocity differs at the top and bottom surface of the cavity. This general case is treated later on in this section. 

\begin{figure}[ht]
\centering\includegraphics[width=0.95\linewidth]{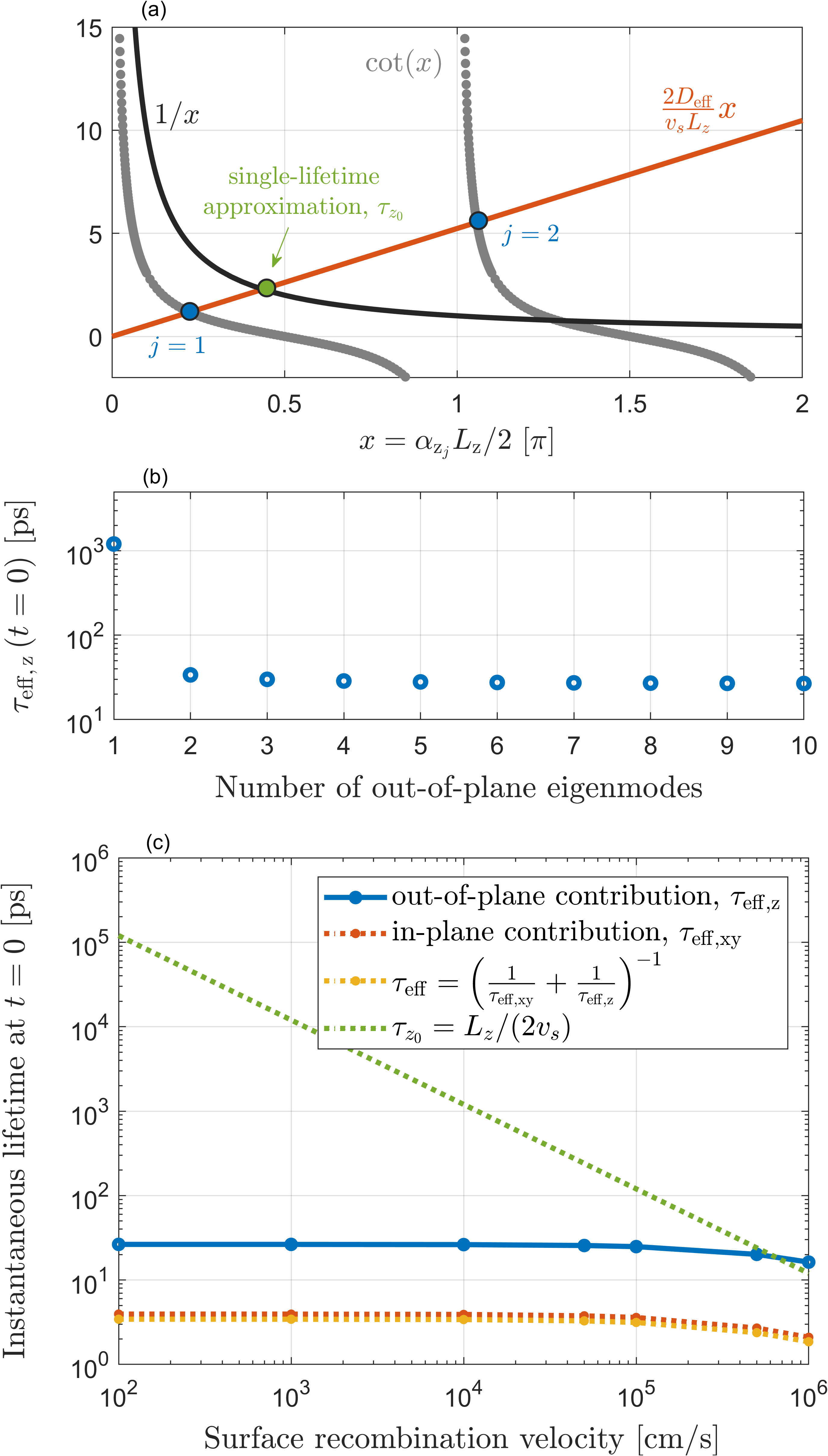}
\caption{\label{fig:sec:SEC5:FIG1} (a) Graphical solution of \Eqref{eq:Sec5:kz} to calculate the out-of-plane diffusion lifetimes. Parameters: $L_{z} = 240\,\mathrm{nm}$ and $v_s = 5\times10^5\,\mathrm{cm/s}$. (b) Out-of-plane instantaneous lifetime at zero time versus the number of eigenmodes taken into account. The EDC ring cavity in \tabref{tab:sec:SEC1:TAB1} is considered with $v_s = 10^4\,\mathrm{cm/s}$. (c) Out-of-plane instantaneous lifetime at zero time (blue) versus the surface recombination velocity. The in-plane instantaneous lifetime (red), the total instantaneous lifetime (yellow), and the out-of-plane single-lifetime approximation (green) are also shown. The same cavity as in (b) is considered.}
\end{figure}
The graphical solution of \Eqref{eq:Sec5:kz} is illustrated in \figref{fig:sec:SEC5:FIG1}a. The intersections (blue bullets) between the straight line (red) on the right-hand side of \Eqref{eq:Sec5:kz} and the cotangent function (grey) on the left-hand side determine the values of $\alpha_{\mathrm{z}_j}$, and hence the diffusion lifetimes. With decreasing cavity thickness and surface recombination velocity, the intersection corresponding to the longest diffusion lifetime moves closer to the $x$-axis origin, where the cotangent function may be approximated with a hyperbola (black). By inserting $\cot{\left(\alpha_{\mathrm{z}_j}L_{\mathrm{z}}/2\right)}\approx\left(\alpha_{\mathrm{z}_j}L_{\mathrm{z}}/2\right)^{-1}$ in \Eqref{eq:Sec5:kz}, one recovers the usual single-lifetime approximation \cite{Nozaki_NatPhot_2010,Yu_OptExpress_2013}  
\begin{equation}
\label{eq:Sec5:Shockley-tauz}
\tau_{\mathrm{z}_0} = \frac{L_{\mathrm{z}}}{2v_s}
\end{equation}
corresponding to the intersection (green bullet) between the straight line and the hyperbola in \figref{fig:sec:SEC5:FIG1}a.

Several eigenmodes generally contribute to the instantaneous lifetime, as already seen for the in-plane diffusion. For example, we consider the EDC ring cavity in \tabref{tab:sec:SEC1:TAB1} with $v_s = 10^4\,\mathrm{cm/s}$. The out-of-plane excitation profile is illustrated in \figref{fig:sec:SEC3:FIG1}d. As shown by \figref{fig:sec:SEC5:FIG1}b, the out-of-plane instantaneous lifetime at zero time significantly differs from the value found by only including the first eigenmode, corresponding to the longest diffusion lifetime. At zero time, the excitation profile dominates the diffusion speed, and higher-order eigenmodes must be considered. However, the convergence with the number of eigenmodes is much faster than previously seen for the in-plane diffusion (cf. \figref{fig:sec:SEC4:FIG4}), and a few eigenmodes are enough for an accurate estimate.

As shown by \figref{fig:sec:SEC5:FIG1}c, the out-of-plane instantaneous lifetime at zero time (blue) is hardly dependent on surface recombination, similarly to the case of in-plane diffusion (cf. \figref{fig:sec:SEC4:FIG6}a). The in-plane instantaneous lifetime (red) and the total instantaneous lifetime (yellow) are also shown. The latter is defined as follows: 
\begin{equation}
\label{eq:Sec5:taueff-def}
\frac{1}{\tau_{\mathrm{eff}}(t)} = -\frac{d\ln\left[h_{\mathrm{eff}}(t)\right]}{dt}
\end{equation}
which leads to
\begin{equation}
\label{eq:Sec5:taueff}
\frac{1}{\tau_{\mathrm{eff}}(t)} = \frac{1}{\tau_{\mathrm{eff,xy}}(t)} + \frac{1}{\tau_{\mathrm{eff,z}}(t)} + \frac{1}{\tau_{\mathrm{bulk}}}  
\end{equation}
based on \Eqref{eq:Sec3:heff-separation}. To focus on the effect of carrier diffusion, we shall assume $\tau_{\mathrm{bulk}}=\infty$. The in-plane diffusion dominates the total instantaneous lifetime. The single-lifetime approximation (green) poorly describes the out-of-plane diffusion, but the error barely affects the total instantaneous lifetime. The other photonic cavities in \figref{fig:sec:SEC1:FIG1-FIG2} feature similar out-of-plane excitation profiles and comparable values of the cavity thickness, leading to similar results (not shown). 

We point out that reducing the cavity thickness accelerates the out-of-plane diffusion, whose contribution to the total instantaneous lifetime may become nonnegligible. However, the change in the out-of-plane excitation profile, which is essential to consider at zero time, should be self-consistently taken into account.   

\begin{figure}[ht]
\centering\includegraphics[width=0.95\linewidth]{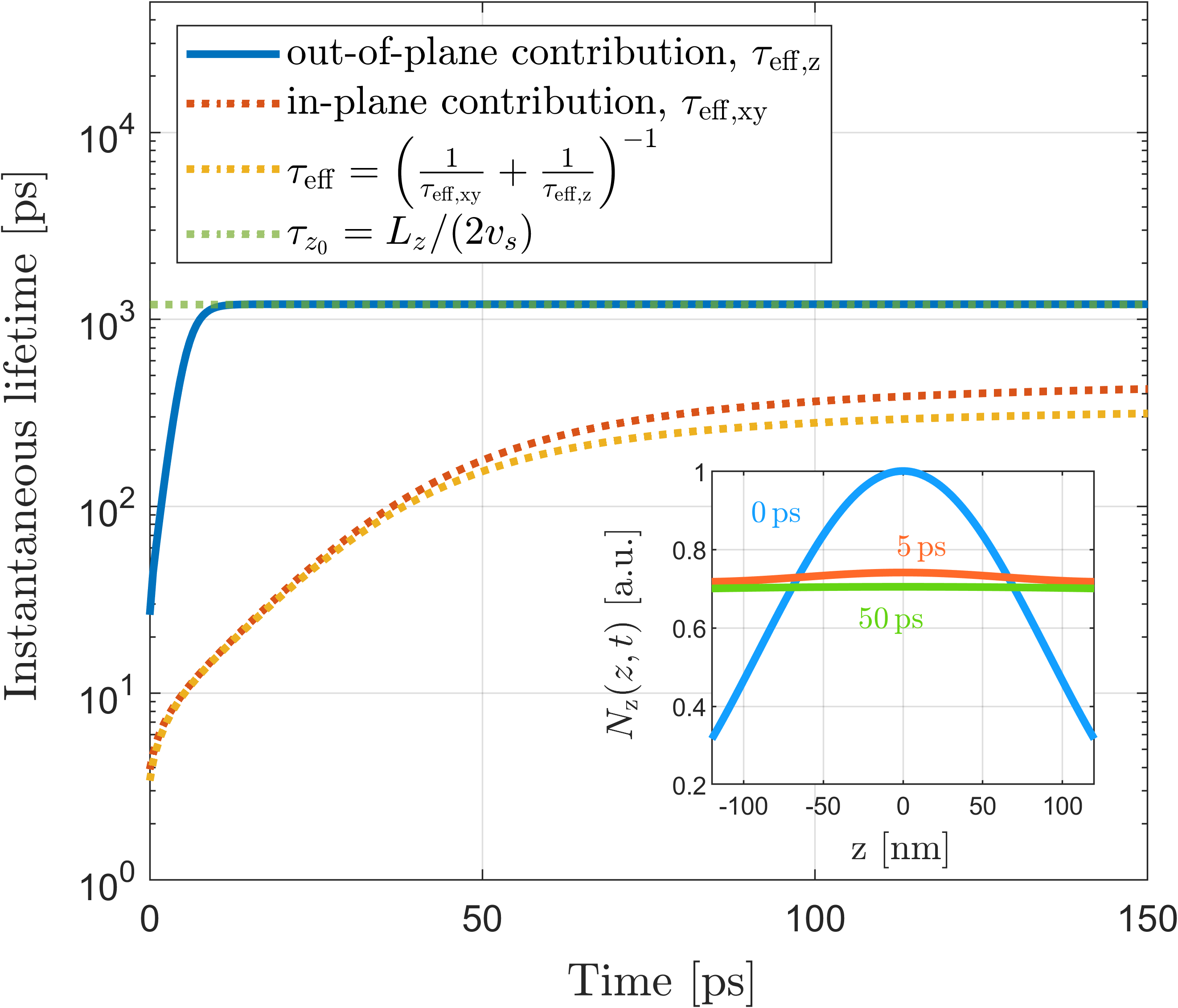}
\caption{\label{fig:sec:SEC5:FIG2} Out-of-plane instantaneous lifetime (blue) versus time. The EDC ring cavity in \tabref{tab:sec:SEC1:TAB1} is considered with $v_s = 10^4\,\mathrm{cm/s}$. The in-plane instantaneous lifetime (red), the total instantaneous lifetime (yellow), and the out-of-plane single-lifetime approximation (green) are also shown. Inset: out-of-plane carrier density distribution at different times versus the out-of-plane position coordinate.}
\end{figure}
The time dependence of the instantaneous lifetimes is illustrated in \figref{fig:sec:SEC5:FIG2}. A surface recombination velocity of $10^4\,\mathrm{cm/s}$ is considered. With increasing time, the out-of-plane lifetime (blue) tends to saturate much faster than the in-plane counterpart (red). The inset displays the out-of-plane carrier density distribution at zero time (light blue), $5\,\mathrm{ps}$ (light red) and $50\,\mathrm{ps}$ (light green) as a function of $z$ (see Sec.$\,$II in the Supplementary Material). The distribution is quickly smeared out, and the carrier decay rate, within a few picoseconds, is only limited by surface recombination. In contrast, the in-plane geometry, which is less trivial, tends to preserve the non-uniformity of the carrier distribution and retard the saturation of the in-plane instantaneous lifetime. The in-plane diffusion is also seen to be faster, and it dominates the total instantaneous lifetime (yellow) at any given time. This is generally the case unless the surface recombination is strong enough.         

\begin{figure}[ht]
\centering\includegraphics[width=0.95\linewidth]{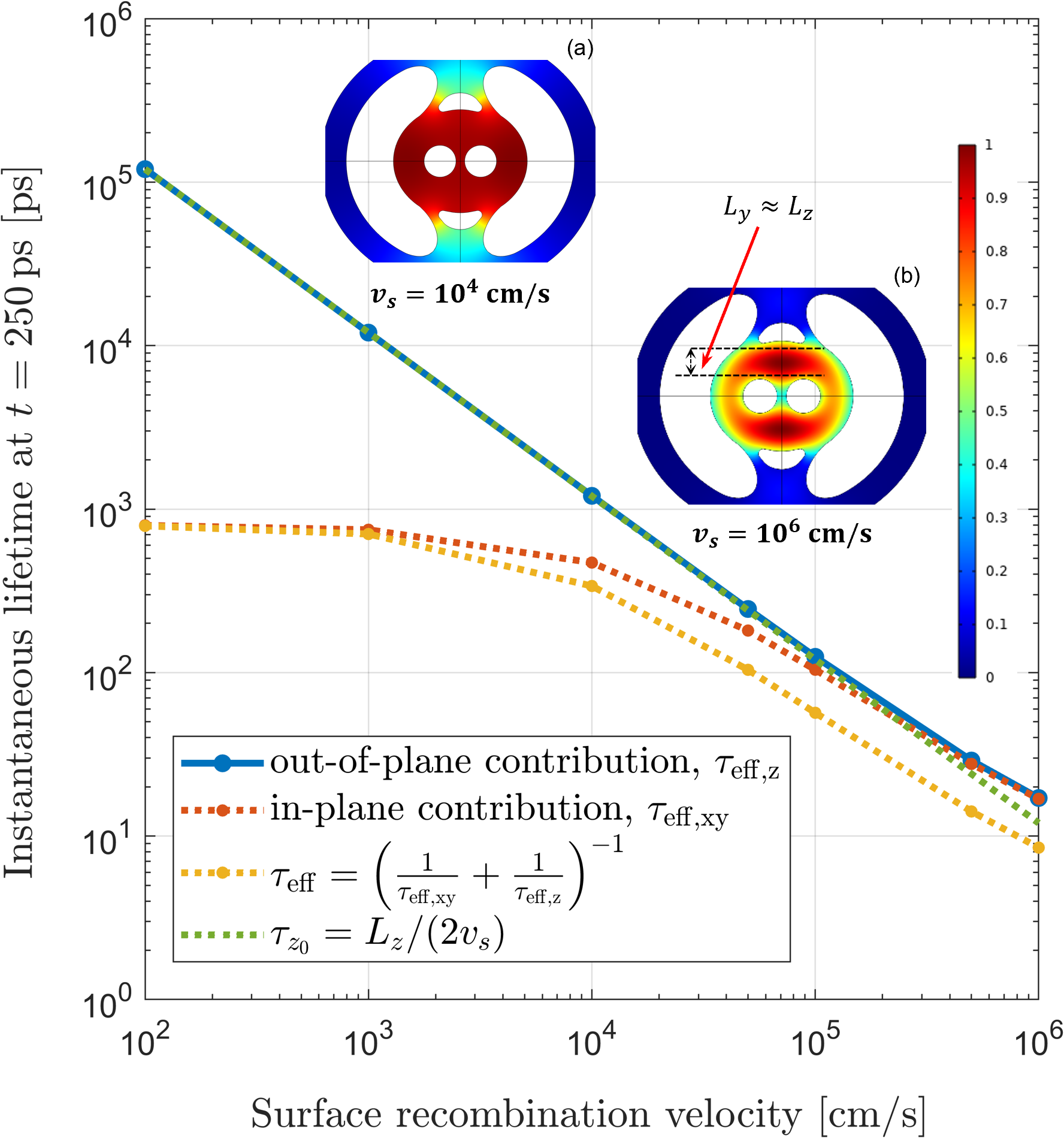}
\caption{\label{fig:sec:SEC5:FIG3} Out-of-plane instantaneous lifetime at $250\,\mathrm{ps}$ (blue) versus the surface recombination velocity. The in-plane instantaneous lifetime (red), the total instantaneous lifetime (yellow), and the out-of-plane single-lifetime approximation (green) are also shown. The EDC ring cavity in \tabref{tab:sec:SEC1:TAB1} is considered with $v_s = 10^4\,\mathrm{cm/s}$. Insets: in-plane carrier density distribution in arbitrary units at $250\,\mathrm{ps}$ for (a) $v_s = 10^4\,\mathrm{cm/s}$ and (b) $v_s = 10^6\,\mathrm{cm/s}$.}
\end{figure} 
\figref{fig:sec:SEC5:FIG3} illustrates the impact of surface recombination on the long timescale, where the in-plane carrier decay rate is close to saturation. For sufficiently large values of surface recombination, the out-of-plane instantaneous lifetime (blue) approaches the in-plane lifetime (red), with a nonnegligible impact on the total lifetime (yellow). The single-lifetime approximation (green) accurately captures the out-of-plane lifetime unless the surface recombination is high. The insets display the in-plane carrier density distribution for (a) moderate and (b) high values of surface recombination. Interestingly, large values of surface recombination tend to split the carriers and trap them in the space between the bowtie holes and the air openings above and below. The size along the $y$-direction, $L_\mathrm{y}$, of this interstitial space (see inset b) roughly corresponds to the cavity thickness, $L_\mathrm{z}$, explaining why the out-of-plane and in-plane lifetimes tend to coincide. We have noticed similar trapping phenomena in the other photonic cavities (not shown). The observation suggests that the in-plane diffusion may be described as a one-dimensional phenomenon at sufficiently long times and high enough values of surface recombination. However, we expect the effective diffusion length to depend on the specific in-plane geometry, which prevents further general considerations.    

If the out-of-plane excitation profile is not symmetric with respect to $z$ or the surface recombination velocity is not the same at the top and bottom surface of the cavity, the out-of-plane diffusion is generally asymmetric. In this case, the values of $\alpha_{\mathrm{z}_j}$ governing the diffusion lifetimes (cf. \Eqref{eq:Sec5:tauz}) are determined by
\begin{equation}
\label{eq:Sec5:eigenval-eq-asymm}
\cot{\left(\alpha_{\mathrm{z}_j}L_{\mathrm{z}}\right)} = \frac{D_{\mathrm{eff}}^2\left(\alpha_{\mathrm{z}_j}L_{\mathrm{z}}\right)^2 - v_{s_t}v_{s_b}L_{\mathrm{z}}^2}{D_{\mathrm{eff}}\,\left(\alpha_{\mathrm{z}_j}L_{\mathrm{z}}\right)\left(v_{s_t} + v_{s_b}\right)L_{\mathrm{z}}}
\end{equation}
with $v_{s_t}$ and $v_{s_b}$ being the surface recombination velocity at the top and bottom surface of the cavity, respectively. The expression of $A_{\mathrm{z}_j}$ (cf. \Eqref{eq:Sec5:Az-cos}) is also generalized (see Sec.$\,$III in the Supplementary Material), and takes into account that the eigenmodes are not even functions of $z$. In practice, the out-of-plane diffusion is asymmetric if the substrate and cladding materials below and above the cavity, respectively, are different (with $F_\mathrm{z}(z)$ thus being asymmetric), or in general if the top and bottom surface feature different defect densities. 

\begin{figure}[ht]
\centering\includegraphics[width=1\linewidth]{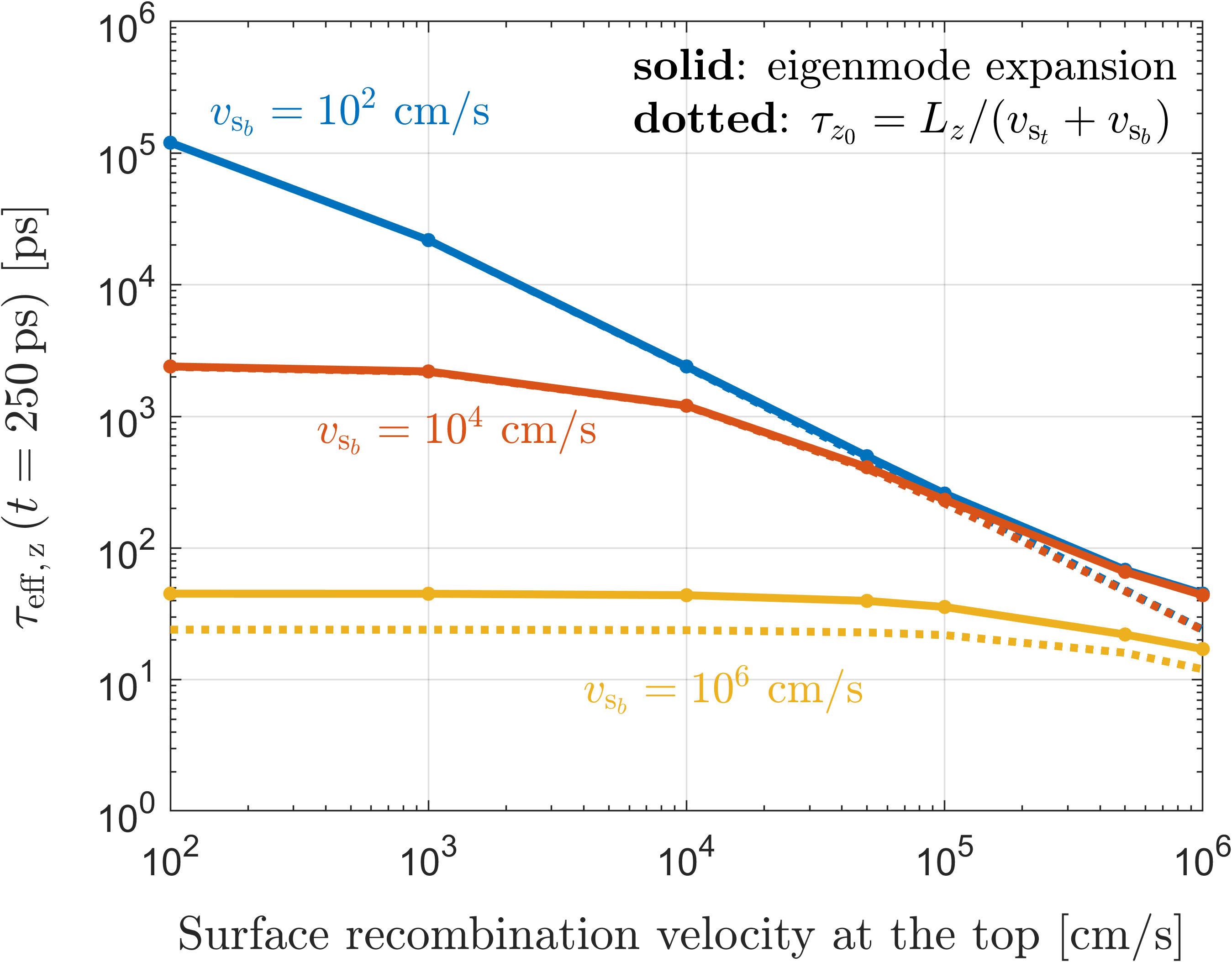}
\caption{\label{fig:sec:SEC5:FIG4} Out-of-plane instantaneous lifetime at $250\,\mathrm{ps}$ (solid) versus the surface recombination velocity at the top surface of the cavity, $v_{s_t}$. The surface recombination velocity at the bottom surface, $v_{s_b}$, is $10^2\,\mathrm{cm/s}$ (blue), $10^4\,\mathrm{cm/s}$ (red) and $10^6\,\mathrm{cm/s}$ (yellow). The single-lifetime approximation (dotted, \Eqref{eq:Sec5:Shockley-tauz}) is also shown with the substitution $v_s\rightarrow\left(v_{s_t}+v_{s_b}\right)/2$. The EDC ring cavity in \tabref{tab:sec:SEC1:TAB1} is considered.}
\end{figure}
For example, we consider again the EDC ring cavity in \tabref{tab:sec:SEC1:TAB1}. The cavity remains $z$-symmetric (with $F_\mathrm{z}(z)$ thus being unchanged), but $v_{s_t}$ and $v_{s_b}$ are now allowed to differ. We focus on the long timescale, where the impact of surface recombination is significant, and the contribution of the out-of-plane diffusion to the total instantaneous lifetime may be nonnegligible. \figref{fig:sec:SEC5:FIG4} shows the out-of-plane instantaneous lifetime at $250\,\mathrm{ps}$ (solid) versus the surface recombination velocity at the top surface of the cavity. We also include the single-lifetime approximation (dotted, \Eqref{eq:Sec5:Shockley-tauz}) with the surface recombination, $v_s$, replaced by the average value of $v_{s_t}$ and $v_{s_b}$. For low (blue) and moderate values (red) of $v_{s_b}$, the out-of-plane instantaneous lifetime as obtained from the eigenmode expansion only differs from the single-lifetime approximation at sufficiently high values of $v_{s_t}$. In contrast, large surface recombination values (yellow) at either the top or bottom surface degrade the accuracy of the single-lifetime approximation, irrespective of the surface recombination velocity at the other surface.              

%% file: sections/section06.tex
\section{Conclusions} \label{sec: conclusions}

In conclusion, we have investigated the diffusion of carriers in semiconductor nanoscale resonators by employing an efficient eigenmode expansion technique. The response function of the mode-averaged (or effective) carrier density is found from the eigenmodes of the ambipolar diffusion equation, at variance with time-domain simulations of previous works \cite{Moille_PRA_2016}. Importantly, we have shown that emerging dielectric cavities with extreme dielectric confinement (EDC) \cite{Hu_ACSPhot_2016,Choi_PRL_2017,Wang_APL_2018,Albrechtsen_NatComm_2022} reduce the time it takes for the carriers to diffuse out of the effective mode area of interest. This is due to the tight confinement of the electric field to a hot spot. Thus, the effective carrier density decay rate is accelerated, which is promising for optical switching applications compared to more conventional geometries. 

The eigenmode approach singles out the contribution to the effective carrier density decay rate due to the in-plane and out-of-plane cavity geometry. The in-plane contribution is found to dominate unless the surface recombination is high and the timescale of interest is sufficiently long. Furthermore, we have quantified the instantaneous decay rate of the effective carrier density, as a function of time and surface recombination, and compared EDC geometries with conventional photonic crystal designs. In contrast, previous works often rely on multi-exponential fits of nonlinear switching experiments \cite{Heuck_APL_2013,Bazin_APL_2014}, where the impact of carrier diffusion is difficult to isolate.       

The eigenmode approach captures the long-timescale decay rate with only a few eigenmodes, making the method advantageous compared to time-domain simulations. Moreover, sorting the eigenmodes by their excitation coefficients significantly improves the method convergence speed on the short timescale. Future works may conveniently exploit the formulation herein illustrated (in short, \Eqref{eq:Sec4:taueffxy:eig}, \Eqref{eq:Sec5:taueffz:eig} and \Eqref{eq:Sec5:taueff}) to systematically optimize the carrier decay rate at given times, e.g., by inverse design \cite{Jensen-Sigmnud_LPR_2011,Molesky_NatPhot_2018}.          